\pdfoutput=1

\documentclass[12pt]{article}

\usepackage{amssymb}
\usepackage{hyperref}
\usepackage{cite}

\usepackage{amsmath}
\usepackage{float}
\usepackage{graphics}
\usepackage{graphicx}

\begin{document}

\begin{titlepage}

\rightline{December 2015}

\vskip 2.2cm

\centerline{\Large \bf  
Plasma dark matter direct detection  
}

\vskip 2.0cm

\centerline{J. D. Clarke and R. Foot\footnote{
E-mail address: 
j.clarke5@pgrad.unimelb.edu.au, rfoot@unimelb.edu.au}}
\vskip 0.7cm

\centerline{\it ARC Centre of Excellence for Particle Physics at the Terascale,}
\centerline{\it School of Physics, University of Melbourne,}
\centerline{\it Victoria 3010 Australia}

\vskip 2.4cm

\noindent
Dark matter in spiral galaxies like the Milky Way may take the form of a dark plasma. 
Hidden sector dark matter charged under an unbroken $U(1)'$ gauge interaction provides a simple and well defined particle physics model
realising this possibility.
The assumed $U(1)'$ neutrality of the Universe then implies (at least) two oppositely charged dark matter components
with self-interactions mediated via a massless ``dark photon'' (the $U(1)'$ gauge boson).
In addition to nuclear recoils
such dark matter can give rise to keV electron recoils
in direct detection experiments.
In this context, the detailed physical properties of the dark matter plasma interacting with the Earth
is required.  This is a complex system, which is here modelled as a
fluid governed by the magnetohydrodynamic equations. These equations are numerically solved for some illustrative 
examples, and implications for direct detection experiments 
discussed. In particular, the analysis presented here leaves open the intriguing possibility that the 
DAMA annual modulation signal is due primarily to electron recoils (or even a combination of electron recoils and nuclear recoils).
The importance of diurnal modulation (in addition to annual modulation) as a means of probing this
kind of dark matter is also emphasised.

\end{titlepage}

\section{Introduction}

Evidence for dark matter in the Universe arises from many sources including: 
large scale structure (e.g.\cite{lss,lss2}), 
the cosmic microwave background (e.g.\cite{cmb,cmb2}), 
and galaxy rotation curves (e.g.\cite{rot}). 
Still, the precise nature of dark matter remains uncertain. 
Collisionless dark matter is a simple and well studied possibility, 
which works very well on large scales, 
but has some shortcomings on galactic scales (e.g. \cite{cusp,bull1,bull2}).
On the other hand, 
it is possible that dark matter has 
a very rich structure. 
This is especially natural if dark matter  
resides in a hidden sector with its own gauge interactions. 
In particular, dark matter might be multicomponent, 
charged under an unbroken dark $U(1)'$ gauge interaction,
i.e. it interacts with itself via a massless ``dark photon''.
It has been suggested that such self-interactions may even
go some way toward ameliorating small scale structure problems (cf. \cite{s1,s2,s3},
see also e.g. \cite{footexplorede} for more recent work).
Mirror dark matter is a theoretically constrained example of 
such a theory (for a review and bibliography see Ref.~\cite{mirror1}) 
and there are many other scenarios 
considered in the literature 
(e.g. \cite{rich1,and2,rich2,rich3,rich4,rich2x,and1,rich5,rich1x,rich6,rich7,rich8}).
In such a framework,
it is possible that the dark matter in the Universe 
exists primarily in a plasma state,
as a macroscopically neutral ``conductive gas'' of
ions with dark charge,
broadly analogous to the state of much of the ordinary matter in the Universe.
It is this ``plasma dark matter'' scenario
that is the subject of this paper.

One important and distinctive property of a multicomponent plasma dark matter halo with light and heavy mass components is the following:
energy equipartition implies light component velocities
which are much larger than those expected under single component virialisation,
and overall $U(1)'$ neutrality implies they can even be much larger than the galactic escape velocity.
It has been pointed out \cite{foot14} that this effect
can give rise to observable keV electron recoils
in direct detection experiments.
It might even be possible to explain the DAMA annual modulation signal \cite{dama1a,dama1b,dama2a,dama2b,dama2c} in this manner,
since the constraints on electron recoils provided by other experiments are generally much weaker than
those of nuclear recoils. 
However, a detailed description of the plasma dark matter density and velocity
distribution in the vicinity of the Earth is required.
This is a highly non-trivial problem.
If the dark plasma has interactions with ordinary matter
then it will by captured by the Earth,
forming an approximate ``dark sphere'' within.
Understanding the interaction of the dark plasma with this 
dark sphere as the Earth moves through the halo 
is therefore of primary importance.
The aim of this paper is to provide a consistent description of this interaction
in order to qualitatively understand the
implications for direct detection.

\begin{figure}
\centering
\includegraphics[width=12.9cm,angle=0]{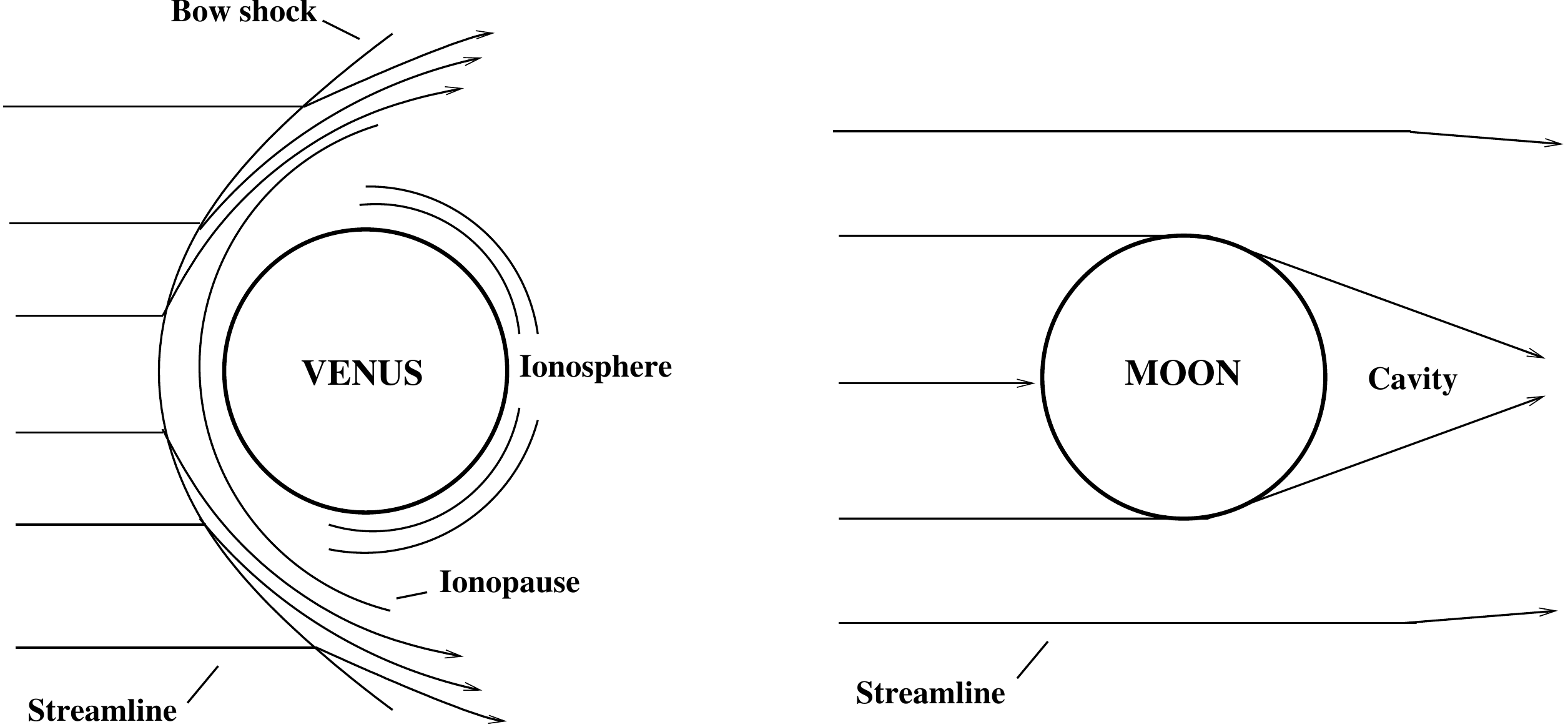}
\caption{{\small
The solar wind interaction with Moon/Venus. These systems appear to represent useful analogues to
the possible ways in which the dark plasma wind interacts with captured dark matter within the Earth.
}}
\label{FigMoonVenus}
\end{figure}

The captured dark sphere within the Earth forms an obstacle to the dark plasma wind.
Two limiting cases can be envisaged:
(1) if the captured dark matter is largely neutral, i.e. poorly conducting, 
then the dark plasma wind will be absorbed by the dark sphere, or;
(2) if the captured dark matter is largely ionised, then the
obstacle forms a conducting sphere which effectively
deflects the dark plasma wind.
Interestingly, these limiting cases appear
analogous to the solar wind interaction with
the Moon and Venus, respectively (e.g. \cite{Spreiter70,Luhmann04});
this is sketched in Figure~\ref{FigMoonVenus}.
The Moon has no magnetic field and no atmosphere, 
so that the solar wind is largely absorbed at the lunar surface
with very little upwind activity.
Venus has no magnetic field, 
but forms an electrically conductive layer at the edge of its ionosphere,
which at first approximation forms an impenetrable obstacle to the solar wind.
These systems have been studied using magnetohydrodynamic (MHD) models,
and this would seem to be an appropriate starting point for studying 
the dark plasma wind interaction with the captured dark matter within the Earth.

This paper is set out as follows. In section 2 we provide a brief introduction to the plasma dark matter model and identify the parameter
space of interest.  In section 3 we discuss some relevant properties of the dark sphere of dark matter captured within the Earth.  
Section 4 examines the dark matter plasma wind interaction with this dark sphere, modelled via magnetohydrodynamics.
In section 5 we consider some general aspects of direct detection of plasma dark matter, identifying the
various sources of event rate modulation. 
Section 6 discusses the  modulation of electron recoils in a specific example.
In section 7 we give some implications of our results
in the light of the current experimental situation, 
and in section 8 our conclusions are drawn.

\section{Plasma dark matter}

Dark matter might reside in a hidden sector with its own gauge interactions.
If the hidden sector contains an unbroken $U(1)'$ gauge interaction, 
then $U(1)'$ neutrality of the Universe implies a 
multicomponent self-interacting dark matter sector
consisting of fermions and/or bosons carrying $U(1)'$ charge.
In the following discussion we consider the minimal two-component case with fermionic dark matter.
The dark matter consists of a ``dark electron'' ($e_d$) and a ``dark proton'' ($p_d$)
with masses $m_{e_d}\le m_{p_d}$ and $U(1)'$ charge ratio
$Z' \equiv |Q'(p_d)/Q'(e_d)|$.
The fundamental interactions are described by the hidden sector Lagrangian:
\begin{eqnarray}
{\cal L} = {\cal L}_{SM} (e, \mu, u, d, A^\mu, ...) \ + \
{\cal L}_{dark} (e_d, p_d, A_d^\mu)  \ + \ {\cal L}_{mix} \ .
\label{1}
\end{eqnarray}
Self-interactions of the dark electron
and dark proton are mediated via the massless dark photon. 
These self-interactions can be defined in terms of 
the $U(1)'$ gauge coupling, $g'$, 
or more conveniently by the dark electron fine structure constant, $\alpha_d \equiv [g'Q'(e_d)]^2/4\pi$.
The dark sector is then fully described by the fundamental parameters: $m_{e_d}, m_{p_d}, Z', \alpha_d$.

One special case of this picture is a thermal relic dark matter scenario
with particle-antiparticle dark matter \cite{rich1}.
\footnote{The
particle-antiparticle case is also special because such dark matter can undergo annihilations into
dark photons. Dark matter annihilations are forbidden in the more general case assuming the minimal particle
content ($A_d, \ e_d, \ p_d$), which 
can be viewed as a consequence of accidental $U(1)'$ dark lepton and dark baryon number 
global symmetries.
}
In this case the parameters are constrained:
$m_{e_d} = m_{p_d} \equiv m_\chi$, $Z' = 1$, and 
$\alpha_d\approx~4\times~10^{-5}~\left( m_\chi/\text{GeV}\right)$.
We will be more interested in the general asymmetric dark matter scenario in which $m_{e_d}\ne m_{p_d}$.
A special case of this is mirror dark matter \cite{mirror1}, 
where the  hidden sector is exactly isomorphic to the standard model
so that an exact discrete $Z_2$ symmetry swapping each ordinary particle with a ``mirror'' particle can be defined \cite{flv}. 
The interactions of the mirror electrons together with the dominant mass component, 
assumed to be mirror helium, is then described by:
$m_{e_d} = m_e \simeq 0.511$~MeV, $m_{p_d} = m_{He} \simeq 3.76$~GeV, $Z' = 2$ and $\alpha_d = \alpha \simeq 1/137$.

The interactions of the dark sector with the standard sector are contained within the
${\cal L}_{mix}$ term in Eq.~(\ref{1}).  
The only renormalisable (and non-gravitational) interaction allowed in the minimal setup is kinetic mixing
of the $U(1)_Y$  and $U(1)'$ gauge bosons \cite{he}, 
which implies also photon - dark photon kinetic mixing:
\begin{eqnarray}
{\cal L}_{mix} = \frac{\epsilon'}{2} \ F^{\mu \nu} F'_{\mu \nu}
\ .
\label{kine}
\end{eqnarray}
Here $F_{\mu \nu}$ and $F'_{\mu \nu}$ denote the field strength tensors for the photon
and dark photon respectively and the dimensionless parameter $\epsilon'$ encodes the strength of the mixing
interaction. The kinetic mixing interaction imbues the dark electron and dark proton
with an ordinary electric charge, proportional to this kinetic mixing parameter, $\epsilon'$ \cite{holdom}.
It is convenient to introduce a new parameter, $\epsilon$, 
such that the magnitude of the dark electron's ordinary
electric charge is $\epsilon e$.
Now, including the dark sector parameters,
the fundamental physics is fully described 
by five parameters: $m_{e_d}, m_{p_d}, Z', \alpha_d, \epsilon$. 


In this paper we are interested in the region of parameter space whereby the dark matter in spiral  
galaxies such as the Milky Way is in the form of a dark plasma.
Typically this requires that
the dark atomic binding energy be much smaller than (or of order)
the temperature of the dark electrons
(see \cite{footexploredb} for more precise calculations). 
The binding energy of the hydrogen-like dark atom 
consisting of a dark proton and a dark electron
is
\begin{eqnarray}
I = \frac{1}{2} \ Z'^2 \alpha_d^2 \mu_d \ ,
\label{isis}
\end{eqnarray}
where $\mu_d = m_{e_d}m_{p_d}/(m_{e_d}+m_{p_d})$ is the reduced mass.
The temperature of the dark electrons is more difficult to determine.
Let us assume for now that the frequency of interactions of the dark electrons and 
dark protons is sufficiently great so that they have approximately the same temperature, 
and that this temperature is approximately the same throughout the halo 
(we will return to this condition shortly).
The halo temperature can then be estimated from the virial theorem \cite{sph}
and also by assuming hydrostatic equilibrium \cite{mirror1}: 
\begin{eqnarray}
T \sim \frac{1}{2} \ \bar m v_{rot}^2 \ ,
\label{3}
\end{eqnarray} 
where $v_{rot}$ is the asymptotic value of the rotational velocity
of the galaxy (for the Milky Way, $v_{rot} \approx 220$ km/s)
and $\bar m$ is the mean mass of the particles in the halo.\footnote{Unless 
otherwise stated we adopt natural units where $\hbar = c = k_B = 1$.}
For a fully ionised halo the mean mass can be determined from $U(1)'$  
neutrality:   
\begin{eqnarray}
\bar m = \frac{m_{p_d} + Z'm_{e_d}}{Z'+1} 
\ .
\end{eqnarray}
The plasma will be fully ionised if $I/T \ll 1$, a condition that reduces to:
\begin{eqnarray}
\frac{I}{T} \simeq 0.20 \ Z'^2 (Z'+1) \left( \frac{\alpha_d}{10^{-2}}\right)^2 \left( \frac{\mu_d}{{\rm MeV}} \right)
\left( \frac{{\rm GeV}}{m_{p_d} + Z'm_{e_d}}\right)\left( \frac{220 {\rm \ km/s}}{v_{rot}}\right)^2 \ll 1 \ .
\label{2x}
\end{eqnarray}

Of course with the 
dark photon massless this type of dark matter is dissipative.  
The plasma halo can radiatively cool via processes such as dark bremsstrahlung and potentially
collapse onto a disk on a timescale less than the Hubble time. 
Thus, there is another condition for such a plasma to exist today, namely that
the cooling timescale is longer than the Hubble time, or that a heating mechanism exists.
The cooling timescale is given in e.g. \cite{rich8} and requiring that this  timescale be longer
than the Hubble time for the Milky Way gives the approximate condition
\begin{eqnarray}
m_{p_d} \gtrsim  20 \left( \frac{{\rm MeV}}{m_{e_d}} \right) \left( \frac{\alpha_d}{10^{-2}}\right)^2 
Z'^{5/3} \  {\rm GeV}
\ .
\label{3x}
\end{eqnarray}
This was derived assuming that cooling is dominated by bremsstrahlung for the most stringent
case of $m_{e_d} \ll m_{p_d}$.
The alternative possibility is that the cooling rate 
is sufficiently high for the halo to have collapsed
but is prevented from doing so due to heating 
\cite{sph,rich8,footexploredb,footexploredc,footexploredd,footexplorede}.
If $\epsilon \sim 10^{-9}$--$10^{-10}$
then sufficient heating of the halo can be provided by ordinary core-collapse supernovae \cite{sph,rich8}.
In that scenario, the halo is viewed as a dynamical object which evolves until an equilibrium configuration
is reached where heating and cooling rates locally balance.

The conditions so far have been derived assuming that interactions were 
sufficiently rapid so that dark electrons and dark protons have
approximately the same temperature.
Let us briefly estimate the parameter space where this assumption is reasonable.
If the mean kinetic energy of the dark electrons happened to be much greater than that of the dark protons,
then the two body Rutherford scattering process
($e_d + p_d \leftrightarrow e_d + p_d$)
would transfer net energy from the dark electrons to dark protons. 
Requiring that the timescale for which dark 
electrons are able to transfer all their excess energy to
the dark protons is less than the Hubble time gives:
\begin{eqnarray}
n_{e_d} n_{p_d} \int \frac{d\sigma}{dE_R} \ E_R v \ dE_R \gtrsim  \frac{T n_{e_d}}{t_H} \ ,
\label{the}
\end{eqnarray}
where $T$ is the temperature of the dark electrons, $t_H \sim 14$ Gyr is the Hubble time and $d\sigma/dE_R$ is the
differential cross section in terms of the recoil energy $E_R$ of the scattered dark proton (approximated as initially at rest
relative to the incoming dark electron of velocity, $v$).  
The cross section for this Rutherford scattering process is given by $d\sigma/dE_R = 2\pi Z'^2\alpha_d^2/(m_{p_d} E_R^2 v^2)$.
Equation~(\ref{the}) can be straightforwardly evaluated with the result depending logarithmically on the integration limits.
The upper integration limit is obtained from kinematics:
$E_R^{max} = 4E_i \mu_d/[m_{e_d} + m_{p_d}]$ (where $E_i \sim T$ is the initial energy
of the dark electron) while the lower integration limit is given in terms 
of the Debye shielding length,
$\lambda_D = \sqrt{T/(4\pi \alpha_d n_{e_d})}$,
the scale over which the dark proton's charge is shielded by the dark electrons.
Assuming $T$ given by the estimate, Eq.~(\ref{3}),
and for typical Milky Way dark matter densities, $\rho_{DM} \sim 0.3\ {\rm GeV/cm^3}$,
we find that Eq.~(\ref{the}) reduces to the condition (for $m_{e_d} \ll m_{p_d}$):
\begin{eqnarray}
m_{p_d} \lesssim  \ 10^2 \ (Z' + 1)^{3/7} \left(\frac{220 \ {\rm km/s}}{v_{rot}}\right)^{6/7} 
\left( \frac{Z' \alpha_d}{10^{-2}} \right)^{4/7} \left( \frac{m_{e_d}}{\rm MeV}\right)^{1/7} 
\label{4x}
\ {\rm GeV\ .}
\end{eqnarray}
Strictly this derivation assumes the case of negligible dissipation and heating of the halo during the Hubble timescale,
and modification is possible in the alternative case.

Equations~(\ref{2x}), (\ref{3x}), and (\ref{4x})
give the rough conditions under which the dark matter is expected 
to take the form of a plasma in galaxies with the mean kinetic energy 
of the dark electrons comparable to that of the dark protons.
It is clear that there is a significant region of parameter space available.
In the limit where the dark electrons are much lighter than the dark protons, 
an important feature emerges:
{\it the velocity dispersion of the dark electrons 
is much greater than that of the dark protons,
and in fact can be even larger than the typical galactic escape velocity}.
The dark electrons are prevented from escaping the galaxy due to $U(1)'$ neutrality;
the plasma is highly conducting, and dark electric forces 
act to keep the plasma neutral over length scales larger than the Debye length.
This is a distinctive feature of the plasma dark matter halo, indeed this behaviour 
is very different from weakly interacting massive particle (WIMP) dark matter
or even a collisional gas of light and heavy neutral components.

Before proceeding further, we conclude this section with a brief discussion relating to the bigger picture.
Cosmological aspects of plasma dark matter have been discussed in the literature in the special
case of mirror dark matter (e.g. \cite{ber1,ignatiev,ber2,foot1,footcmb,paolo1z,mirror1}). There is also a growing literature exploring  
more generic models with dark matter featuring unbroken $U(1)'$ (dark photon) gauge interactions (e.g.
\cite{rich1,and2,rich2,rich4,rich2x,rich5,rich1x,rich6,rich7,rich8,v1,v2,cyr2,cyn,vogel}).
These studies demonstrate, among other things, that 
this type of self-interacting dark matter 
can reproduce the success of collisionless cold dark matter on very large
scales with deviations expected on smaller scales. 
Indeed, the self-interactions might be important in
addressing long-standing problems on small scales, 
including an explanation for cored dark matter profiles within galaxies.
However, there are also upper bounds on the strength of such self-interactions.
In particular, merging cluster systems have been used 
as a probe of dark matter self-interactions, the Bullet cluster
system being one well studied example \cite{bullet1,bullet2}.
Attempting to evaluate robust bounds on plasma dark matter from 
cluster collisions is, unfortunately, non-trivial. 
Firstly, the plasma dark matter self-interactions on cluster scales require careful modelling, 
as it is the collective plasma effects (i.e. not hard collisions) which potentially dominate \cite{fin6}.
Secondly, the plasma dark matter distribution
within the cluster is required but is very uncertain.
If the dark matter is sufficiently ``clumpy''
then the dark matter associated with each cluster 
can pass through each other essentially unimpeded, 
thereby consistent with the observations \cite{sil6}.
This depends on the fraction of the dark matter bound
to individual galaxy halos compared to the diffuse cluster component,
which is difficult to determine as it depends on the
detailed properties of the cluster and its history,
as well as the properties of the dark plasma.
In short, despite the fact that such cluster mergers do in principle constrain
the plasma dark matter parameters, it is not yet possible to write down any 
reliable and caveat-free bounds.

\section{Dark matter within the Earth}

The physical properties of the halo dark matter in the vicinity of the Earth
are influenced by the way in which the halo dark matter interacts with 
the dark matter bound within the Earth.
It is therefore pertinent to try to understand some of the relevant features 
of this ``dark sphere'' of influence.
Some aspects of this problem have already been discussed 
for the specific cases of mirror dark matter \cite{fdiurnal} and
for more generic dissipative dark matter models \cite{sunny}. 
The discussion below draws on this work 
and extends it to the more general plasma case.

How does dark matter within the Earth arise?
The kinetic mixing induced interaction with standard matter will 
occasionally trap some halo dark matter particles
within the Earth during its formation phase and subsequently.
Eventually sufficient dark matter accumulates so that
further dark matter capture is primarily facilitated by self-interactions
of halo dark matter with this captured dark matter.
Once captured, it is expected to quickly thermalise
with the ordinary matter within the Earth
via the kinetic mixing interactions,
to a temperature $T_E \sim 5000$~K (0.4~eV).
If $m_{p_d}\gg $~MeV, 
this is much cooler than the halo temperature and the 
dark protons and dark electrons can potentially combine into neutral dark atoms.
The dark sphere will be largely neutral (ionised) 
if $T_E \ll I$ ($T_E \gg I$), 
where $I$ is the dark atomic binding energy given already in Eq.~(\ref{isis}).
This motivates two limiting cases:
a neutral ``Moon-like'' case in which the dark sphere largely absorbs the dark plasma wind, and;
an ionised ``Venus-like'' case in which the dark sphere largely deflects the dark plasma wind
by way of a current-carrying sheet at the ``dark ionopause''
(located where the plasma wind and ``dark ionosphere'' pressures equilibrate).
In addition to its ionisation state,
the other defining feature of the dark sphere is its effective size. 
Let us define a parameter, $R_{DM}$,
which corresponds to the dark plasma wind stopping radius for the Moon-like case,
and the dark ionopause radius for the Venus-like case.
We will now attempt to estimate $R_{DM}$ 
in terms of the fundamental plasma parameters.

If the dark sphere is Moon-like,
then the (relatively stationary) dark protons
accumulate at the ``geometric'' rate
\begin{eqnarray}
\frac{dN_{p_d}}{dt} \approx \pi R_{DM}^2 v_{rot} n_{p_d}
\ . 
\label{1z}
\end{eqnarray}
Dark electrons will be captured at a similar rate: $dN_{e_d}/dt = Z'dN_{p_d}/dt$ given 
the expected approximate $U(1)'$ charge neutrality of the Earth.\footnote{This
corresponds to a captured mass of $\sim 10^{15}(R_{DM}/R_E)^2$~kg
if $R_{DM}$ has remained roughly constant throughout Earth's history;
note that this is much smaller than $M_E \sim 10^{24}$~kg.}
This represents an upper bound for the accumulation rate
in the Venus-like case,
though it might still be a useful estimate
so long as a significant fraction ($\gtrsim 1$\%) 
of the halo wind is stopped within the Earth.
Loss rates due to mechanisms such as 
thermal escape and dark atmospheric stripping
are difficult to evaluate.
Naturally, any estimate of the total
amount of dark matter captured within the Earth is uncertain.
Fortunately, it turns out that $R_{DM}$
depends only weakly on the total number of Earth bound dark matter particles.

If we equate the radial temperature profile of the dark sphere gas/plasma
with that of the Earth \cite{PREM},
the dark thermal pressure is given by
$p(r) = \rho_{DM} (r) T_E (r)/\bar m$. Here $\bar m$ is the mean mass taking into account 
the ionisation state of the captured dark matter at the temperature $T_E(r)$. 
Assuming $m_{p_d} \gg m_{e_d}$, then $p(r) \simeq \xi \rho_{DM} (r) T_E (r)/m_{p_d}$
where $\xi = 1$ ($\xi = Z'+1$) for the Moon-like (Venus-like) case.
The mass density profile $\rho_{DM} (r)$ of the captured dark matter
can then be estimated from the hydrostatic equilibrium condition (with spherical symmetry assumed):
\begin{eqnarray}
\frac{dp(r)}{dr} = -\rho_{DM} (r) g(r)
\ ,
\end{eqnarray}
where $g(r) \simeq G_N \int_0^r \rho_E 4\pi r'^2 dr'/r^2 $
is the local gravitational acceleration within the Earth,
almost entirely due to the ordinary matter component.
Numerical work \cite{fdiurnal,sunny} indicates that
$\rho_{DM}(r)$ falls exponentially, with a scale length 
inversely proportional to the square root of $m_{p_d}$. 
This behaviour can be understood via simple analytical considerations.  
For radially constant $T, \rho_E$, and assuming $m_{p_d} \gg m_{e_d}$, 
the hydrostatic equilibrium condition has the analytic solution:
\begin{eqnarray}
\rho_{DM} (r) = \rho_{DM} (0) \ e^{-r^2/R_h^2} \ ,
\end{eqnarray}
where 
\begin{eqnarray}
R_h &= &  \left( \frac{3T_E \xi}{G_N \rho_E 2 \pi m_{p_d}} \right)^\frac12
\nonumber \\
\Rightarrow \frac{R_E}{R_h} &\simeq & 1.2\ 
 \left( 5000 \text{ K}\over T_E \right)^\frac12 
 \left(\frac{\rho_E}{10 \ {\rm g/cm}^3}\right)^\frac12 
 \left(\frac{m_{p_d}/\xi}{{\rm GeV}}\right)^\frac12
\ .
\end{eqnarray} 
Evidently the dark matter density profile 
depends only on the mass and ionisation state of the dark matter particles.
Here, $R_h$ is the dark sphere scale length, which can be viewed as a rough estimate for $R_{DM}$.
If $m_{p_d}\gtrsim$~few GeV, then $R_h$ is expected to be within the Earth.
In the alternative case which suggests $R_E/R_h<1$,
the thermal equilibrium assumption used in this calculation breaks down,
and we would expect thermal escape and dark atmospheric stripping effects
to act to keep $R_E/R_{DM}\gtrsim 1$,
though it is difficult to say much more than this without detailed calculations.

For the Venus-like case, $R_h$ can only provide a rough estimate
for the location of the dark ionopause,
as the ram pressure of the dark plasma wind
and the pressure of the captured dark sphere
can typically vary by many orders of magnitude.
For the Moon-like case
the stopping radius $R_{DM}$ scales with $R_h$ 
but also depends on the dark matter self-interaction cross section 
and hence on the other fundamental parameters.
In fact, the $R_{DM}$ value for dark electrons 
is not the same as that for
dark protons as their self-interaction cross sections are different.
For now, we shall ignore this subtlety and focus on the $R_{DM}$ scale relevant for dark protons.
Explicit calculations \cite{sunny} that take into account the Earth's temperature and density
profiles indicate that $R_{DM}$ for dark protons is roughly:
\begin{eqnarray}
\frac{R_E}{R_{DM}} \approx \left( \frac{10^{-2}}{\alpha_d} \right)^{0.06} \ 
\left( \frac{m_{p_d}/\xi}{5\ {\rm GeV}} \right)^{0.55}  \ 
\left( \frac{1}{Z'}\right)^{0.14} \ ,
\end{eqnarray} 
which is approximately  valid for $5 \times 10^{-4} \lesssim \alpha_d 
\lesssim 5\times 10^{-2}$, 5 GeV $\lesssim m_{p_d}/\xi \lesssim$ 300 GeV, $1 \lesssim Z' \lesssim 40$.
Again, if the plasma parameters suggested $R_E/R_{DM}<1$, 
we expect that dark sphere interactions with the dark plasma wind will keep $R_E/R_{DM}\gtrsim 1$.

In the next section we proceed to describe the interaction of the halo dark matter with the dark sphere. 
We model the halo dark matter using the MHD equations.
Specifically we shall consider single fluid equations describing the system
in terms of the total density, temperature, bulk velocity, and dark magnetic field.
Such a description is only valid over distance scales
larger than the Debye length: $\lambda_D = \sqrt{T/(4\pi \alpha_d n_{e_d})}$.
For the dark plasma near the Earth, 
i.e. for $v_{rot} \approx 220$ km/s and $\rho_{DM} \approx 0.3 \ {\rm GeV/cm^3}$,
\begin{eqnarray}
\lambda_D \sim 0.2 \ [Z'(Z'+1)]^{-1/2}\left( \frac{m_{p_d} + Z' m_{e_d} }{{\rm GeV}}\right)\left(\frac{10^{-2}}{\alpha_d}\right)^\frac12
\ {\rm km}.
\end{eqnarray}
Requiring  $\lambda_D/R_{DM} \ll 1$ imposes only a very mild restriction on parameter space. 


\section{Dark plasma wind and near-Earth environment}

The interaction of the dark plasma wind with a 
macroscopic obstacle (length scale $\gg \lambda_D$) 
may be modelled via the magnetohydrodynamic (MHD) equations.
It is a remarkable fact that the MHD equations
can be derived as the moment equations of dark ion 
distribution functions obeying the kinetic Vlasov equations
for a collisionless plasma;
thus, even in the absence of hard collisions,
collective effects of the long-range Coulomb force
give rise to a fluid-like behaviour.
For a perfectly conducting ideal fluid, 
the MHD equations take the form (in cgs units)
\begin{align}
\frac{\partial \rho}{\partial t} + \nabla  \cdot (\rho {\bf v}) =&\ 0,
\nonumber \\
\frac{\partial (\rho {\bf v})}{\partial t} 
+ \nabla \cdot \left[ \rho {\bf v} {\bf v} + {\bf I}\left(p + \frac{B^2}
{2}\right) - {\bf B}{\bf B} \right] =&\ 0,
\nonumber \\
\frac{\partial E}{\partial t} + \nabla \cdot \left[
\left( E + p + \frac{B^2}{2} \right){\bf v} - {\bf B} \left({\bf v} \cdot {\bf B}\right)
\right] =&\ 0,
\nonumber \\
\frac{\partial {\bf B}}{\partial t} + \nabla \cdot \left({\bf v}{\bf B}-{\bf B}{\bf v}\right) =&\ 0,\label{EqMHD}
\end{align}
where $\rho$ is the mass density,
$p$ is the thermal pressure,
${\bf v}$ is the bulk velocity, 
${\bf B}$ is the (dark) magnetic field (a factor of $1/\sqrt{4\pi}$ has been absorbed),
and $E= \rho v^2/2 + B^2/2 + p/(\gamma - 1)$ is the energy density,
where we take an ideal gas equation of state with 
a ratio of specific heats $\gamma=5/3$.

In practice the MHD equations are solved in a dimensionless form by setting 
$\tilde{\rho} = \rho/\rho_0$, 
$\tilde{L} = L/L_0$, 
$\tilde{{\bf v}} = {\bf v}/v_0$,
$\tilde{p} = p/(\rho_0 v_0^2)$, 
$\tilde{t} = t/(L_0/v_0)$, 
$\tilde{{\bf B}} = {\bf B}/\sqrt{4\pi\rho_0 v_0^2}$.
For the dark plasma wind it is convenient to take
$\rho_0 = 0.3$~GeV/cm$^3$, $L_0 = R_{DM}$, 
and $v_0=c_s$, where $c_s$ is the sound speed in the plasma far from the Earth ($r \gg R_E$),
\begin{align}
 c_s = \sqrt{\frac{\gamma p}{\rho}} = \sqrt{\frac{\gamma T}{\bar{m}}} 
   \sim \sqrt{\frac{\gamma}{2}} v_{rot}.
\label{sound4}
\end{align}
Once these dark plasma units are set,
the (quasi-)stable steady state solutions we are interested in
will only depend on the wind mach number $M=v_\infty/c_s$
and the magnetic field strength $\tilde{B}_\infty$ far from the Earth.
The quantity $v_\infty$ is the
plasma wind speed (as measured in the Earth frame) far from the Earth, which is a time-dependent quantity 
due to the Earth's orbital motion: 
\begin{eqnarray}
v_\infty 
= v_{\odot} + \Delta v_E \cos \omega (t - t_0) \label{EqvE}
\end{eqnarray}
where $\omega = 2\pi/{\rm year}$, $v_{\odot} = v_{rot} + 12$ km/s (the 12 km/s correction is due to the Sun's peculiar velocity) 
and $\Delta v_E \simeq 15$ km/s results from the Earth's orbital motion.
Evidently, $v_\infty$ varies by $\pm \Delta v_E$ during the year with a maximum at $t = t_0 \simeq 153$ days (June 2$^{nd}$).


As suggested by the tilde in Eq.~(\ref{sound4}), the local sound speed 
is not known precisely [cf. the temperature Eq.~(\ref{3})].
It is worth remarking here that the phenomenology
is rather sensitive to the value of $c_s$ that is realised.
This is because $c_s$ lies very close to the
plasma wind speed $v_\infty$,
so that the Mach number straddles $M\sim 1$
throughout the year.
Three distinct regimes can immediately be identified: 
the supersonic regime $c_s\lesssim v_\odot - \Delta v_E$;
the subsonic regime $c_s\gtrsim v_\odot + \Delta v_E$, and;
the intermediate regime $v_\odot - \Delta v_E \lesssim c_s \lesssim v_\odot + \Delta v_E $.
In order to explore a representative range of possibilities in these models
we choose to study Mach numbers $M\approx 0.74$--$1.77$
(i.e.  $c_s=140$--$290$~km/s for $v_{rot} = 220$ km/s).

As discussed in the previous section,
if the plasma dark matter has some interaction(s) with the standard matter 
then it will be captured within the Earth,
forming an approximate ``dark sphere'' of dark protons and dark electrons
which may or may not have recombined into dark atoms.
For simulations we consider two limiting cases:
\begin{enumerate}
 \item ``Moon-like'': the large majority 
 of the captured dark plasma is in the form of dark atoms,
 and therefore cannot carry a significant dark current.
 In this case, to first approximation, 
 the dark sphere acts as a perfect absorber of the dark plasma wind,
 much like the Moon in the solar wind.
 \item ``Venus-like'': if a sufficient proportion of the captured dark plasma is ionised,
 then an ionospheric surface layer exists on the dark sphere.
 A current-carrying sheet then forms at the ionopause and,
 to first approximation, the dark sphere acts as a perfect spherical conductor which 
 deflects all of the dark plasma wind, much like Venus
 in the solar wind.
\end{enumerate}
We emphasise that these are first approximations of limiting cases.
Satellite experiments have shown that
MHD simulations employing these approximations 
give good descriptions of the Moon \cite{Cui08,LiangHai13} 
and Venus \cite{Tanaka93,Cable95,DeZeeuw96,Kallio98} solar wind systems,
and we adapt them here as well-motivated paradigm cases
in order to gain useful insight.\footnote{In
the Moon-like limiting case, the dark sphere consists predominately of neutral dark atoms, that is,
a poorly conducting medium. 
Since the dark electron and dark proton stopping
distances within the Earth are in general not equal due to their differing interaction cross sections in the Earth
frame,  significant dark charge
separation within the Earth is possible and hence current flows. While dark electric fields are not expected to directly influence
the halo dark matter distribution at the Earth's 
surface ($r=R_E$) where conductivity is expected to be high and effective Debye screening should
occur, there remains
the possibility that dark magnetic fields generated due to the current flows could have important
implications for the distribution of halo dark matter at the Earth's surface.
Unfortunately, such effects are very difficult to estimate, and no attempt to model them has been undertaken here.
}

We solve the MHD equations numerically within the
\textsc{Pluto v4.2} simulation framework \cite{PLUTO}
utilising \textsc{Chombo v3.2} \cite{CHOMBO} for adaptive mesh refinement.\footnote{We 
make the relevant code and some example datasets 
publicly available at \url{http://github.com/jdclarke5/DarkSphere};
see the \texttt{readme.md} file therein for more information.}
The coordinate system is defined in the frame of the dark sphere,
with the origin at the dark sphere centre and
the ${\bf z}$ axis pointing in the wind direction.
We consider the $yz$ plane (assuming azimuthal symmetry)
on a polar 2048$\times$2048 equivalent grid with spatial extent 
$1\le \tilde{r} \le 12$ and $0\le \theta \le \pi$.
Simulations were performed with inflowing dark plasma 
Mach numbers $M=0.74$--$1.77$ (in steps of 0.01)
and $B=0$.
These unmagnetised simulations are relevant when the
thermal pressure dominates over the magnetic pressure,
i.e. when the plasma beta $\beta = p/(B^2/2) \gg 1$.
Far from the Earth, $\beta = 2c_s^2\rho_0/(\gamma B^2)$,
implying the rough requirement $\tilde{B}_\infty \ll 1$.
For $\alpha_d=\alpha$ this translates to
$B_\infty \ll 5 \ [c_s/(200\text{ km/s})]$~nT,
to be compared with typical values within 
the galactic (intergalactic) medium of $\sim 0.1$--10 nT ($\sim 0.1$~nT).
Note that the existence of a significant magnetic field
will generally break the azimuthal symmetry of the system
and potentially change the phenomenology appreciably.
Some tests showed that our results are valid for 
the field-aligned case with ${\bf \tilde{B}_\infty}\lesssim 0.6 \ {\bf z}$,
after which we saw a sharp change in the system's behaviour.
Study of the magnetised case is left for potential future work.

All simulations are initialised with a flat density
and are allowed to evolve to a steady state.
The dark plasma wind inflows at the $\tilde{r}=12$ boundary 
when $\theta > \pi/2$, and outflows when $\theta < \pi/2$.
In the Moon-like case we take the $\tilde{r}=1$ surface boundary condition
as absorbing ($v_r \le 0$) for the windward side and
reflective ($v_r = 0$) for the leeward side.
This boundary condition was adopted for Moon simulations in previous works \cite{Holmstrom1202,LiangHai13}.
For the Venus-like case the $\tilde{r}=1$ surface boundary condition 
is fully reflective ($v_r=0$).
Our simulations were validated against the
Moon and Venus simulations of Refs.~\cite{Holmstrom1202,LiangHai13,DeZeeuw96}
for solar wind parameters ($n_p\approx 10\text{ cm}^{-3}, M\approx 7, \tilde{B}\approx 1$).

\begin{figure}
 \makebox[\textwidth][c]{
 \includegraphics[width=1.2\textwidth]{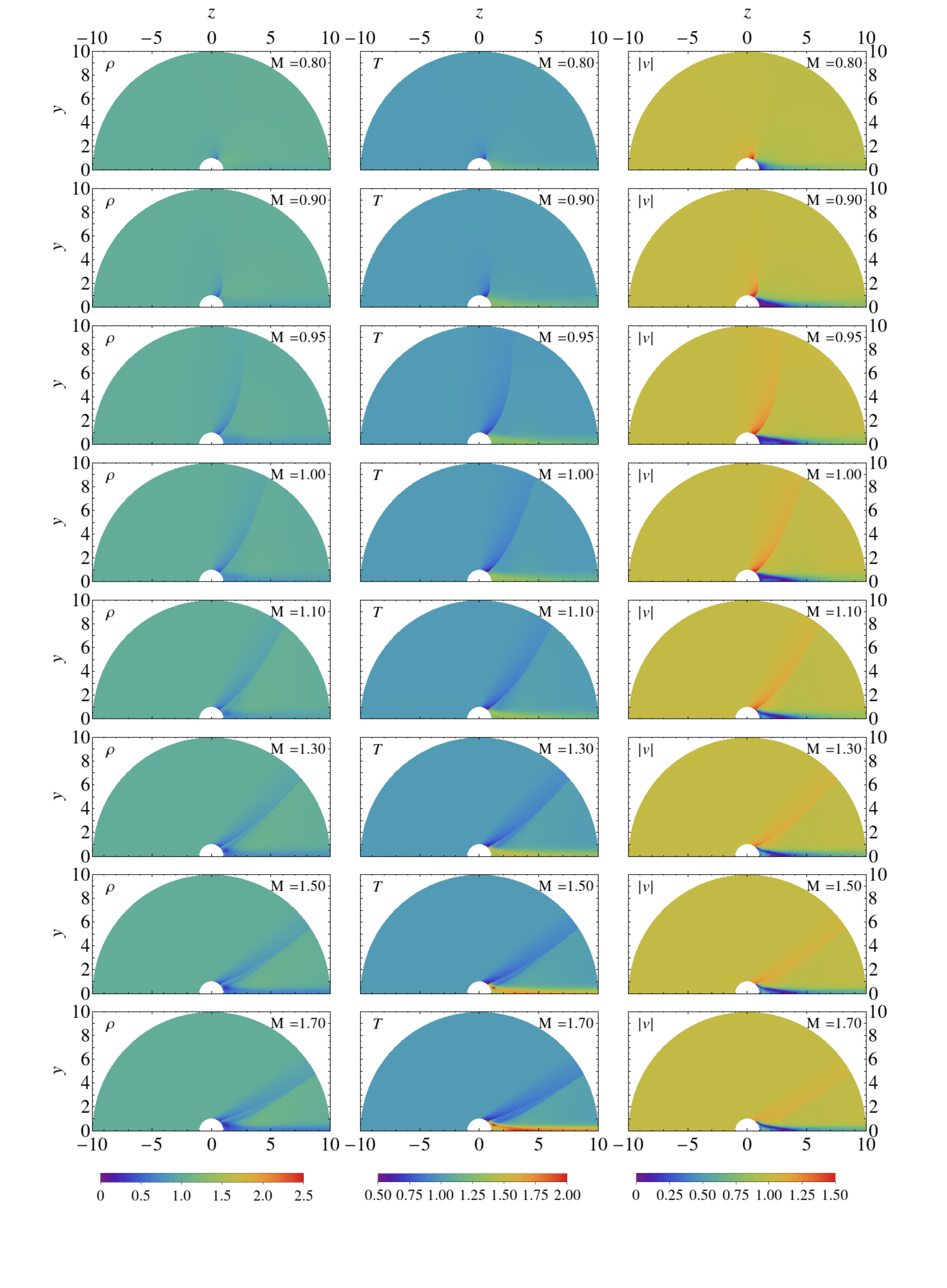}}
 \vspace{-2cm}
 \caption{{\small Moon-like ($B=0$): 
 normalised density, temperature, and absolute velocity solutions for 
 various Mach numbers.
 }}
\label{FigMoonB0}
\end{figure}

\begin{figure}
 \makebox[\textwidth][c]{
 \includegraphics[width=1.2\textwidth]{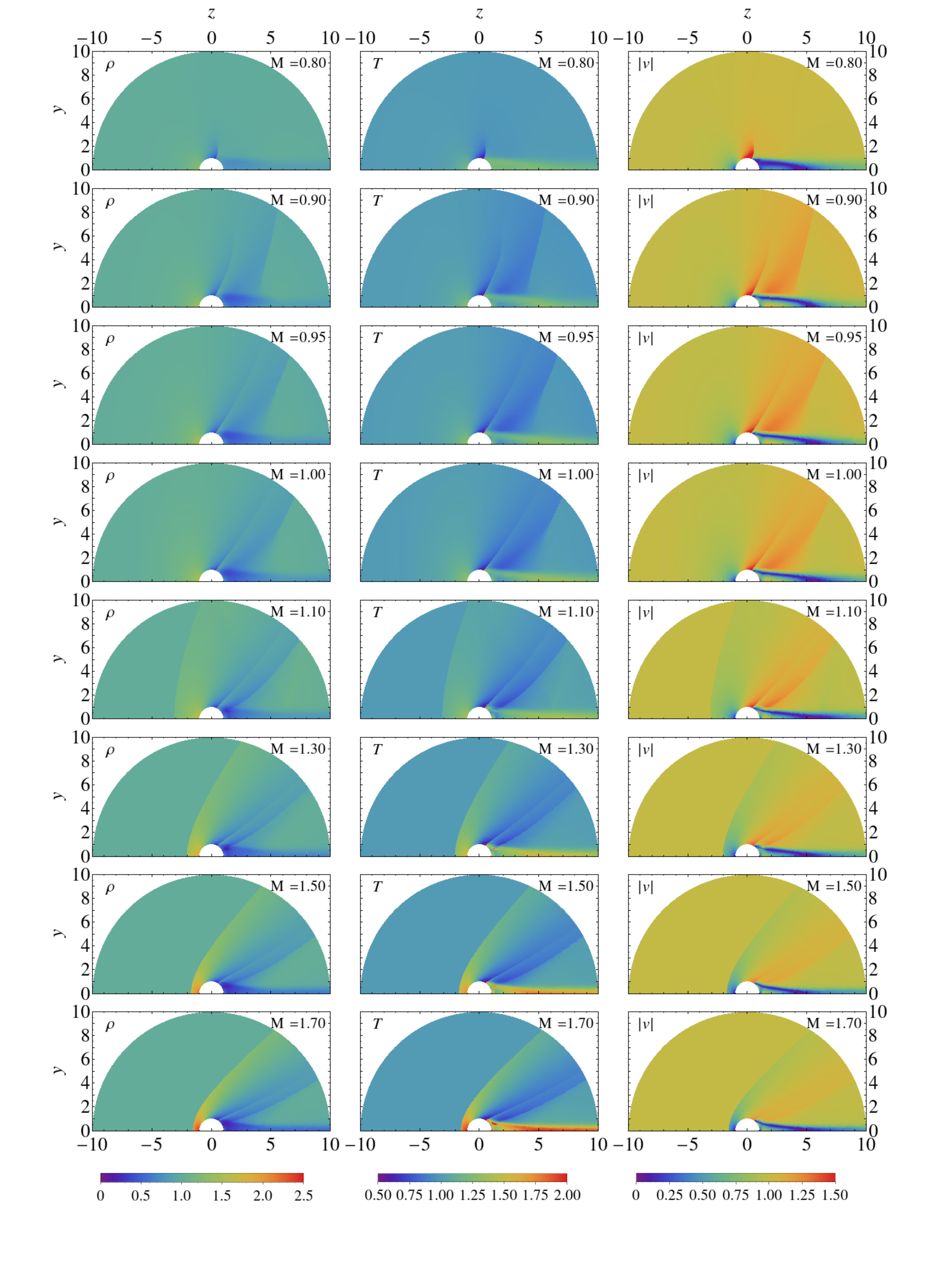}}
 \vspace{-2cm}
 \caption{{\small Venus-like ($B=0$): 
 normalised density, temperature, and absolute velocity solutions for 
 various Mach numbers.
 }}
\label{FigVenusB0}
\end{figure}

The solutions for a collection of Mach numbers are shown
in Figures~\ref{FigMoonB0} and \ref{FigVenusB0}.
We show the distributions for the 
density $\rho$, 
temperature $T=\bar{m}p/\rho$,
and absolute velocity $|v|$,
normalised to their values far from the Earth.
The solutions are characterised by the
existence of various shocks (i.e. abrupt discontinuities),
where the local bulk velocity exceeds the local sound speed.
There are three dominant features:
the downwind wake region of underdense hot plasma;
the tail shock, which detaches from the sphere just below $\theta=\pi/2$
and recedes with increasing Mach number, and;
the upwind bow shock, which is only present in the Venus-like case
(there is very little upwind activity in the Moon-like case),
defining the edge of an induced magnetosphere \cite{Spreiter70,Luhmann04}.

These distributions illustrate the non-trivial (and time-dependent) 
dark plasma environment which surrounds the captured dark sphere 
and encompasses the Earth.
It is clear that there are implications for
direct detection experiments,
and we will discuss these presently.


\section{Plasma dark matter direct detection: General considerations}

Plasma dark matter can potentially be probed by both nuclear and electron recoils.
Kinematic considerations suggest that
dark electrons scattering off electrons (dark protons scattering off nuclei) would be of relevance
if $m_{e_d} \sim m_e$ ($m_{p_d} \sim 10$--100~GeV).
As already emphasised, plasma dark matter has the distinctive feature that
the dark electrons and the dark protons have comparable kinetic energies given by Eq.~(\ref{3}).
Consequently, if the dark proton is sufficiently heavy ($\gtrsim$ GeV), 
electron recoils in the keV range are possible, 
making dark matter scattering off electrons 
(in addition to nuclear recoils)
an important means of probing this type of dark matter.

The rate of dark matter interactions in a direct detection experiment
depends on both the properties (density and velocity distribution) of the dark matter 
particles and the interaction cross section.
Let us define the dark electron and dark proton
velocity probability density functions as
$f_{e_d}({\bf v})$ and $f_{p_d}({\bf v})$
in the Earth frame.
It is generally expected that these functions are both space and time dependent.
The time dependence arises from the 
velocity of the Earth with respect to the dark matter halo, ${\bf v}_E(t)$,
given by Eq.~(\ref{EqvE}).
To make this dependence explicit we rewrite 
$f_{e_d}({\bf v}) \to f_{e_d}({\bf v}; {\bf x}, {\bf v}_E(t) )$ and 
$f_{p_d}({\bf v}) \to f_{p_d}({\bf v}; {\bf x}, {\bf v}_E(t) )$.
The local number densities are also generally space and time dependent:
$n_{e_d} \to n_{e_d}({\bf x}, {\bf v}_E(t) )$,
$n_{p_d} \to n_{p_d}({\bf x}, {\bf v}_E(t) )$.
The {\it local} differential interaction rate
(i.e. at some point, ${\bf x}$, in space near the Earth)
for dark electron scattering off electrons is then:
\begin{eqnarray}
\frac{dR_e}{dE_R}({\bf x}, t) = 
N_e  n_{e_d}({\bf x}, {\bf v}_E(t))
\int^{\infty}_{|{\bf{v}}| > v_{min}
}
\frac{d\sigma}{dE_R}
f_{e_d}({\bf v};{\bf x},{\bf v}_E(t) ) \ |{\bf{v}}| \ d^3v \ ,
\label{f3}
\end{eqnarray}
where 
$E_R$ is the recoil energy of the target particle (electron).
Also, 
$N_e$ is the number of target electrons in the detector,
$d\sigma/dE_R$ the relevant cross section,
and the lower velocity limit $v_{min} (E_R)$ 
is given by the kinematic relation
\begin{eqnarray}
v_{min} = \frac{\sqrt{m_e E_R/2}}{\mu} 
\label{v}
\end{eqnarray}
with $\mu = m_e m_{e_d}/(m_e+m_{e_d})$ the reduced mass
(the target electrons are approximated as being at rest).
The rate of dark proton - nuclei scattering has a similar form.

What the detector will actually measure is the rate Eq.~(\ref{f3})
time-averaged over its position ${\bf x}(t)$ in space.
In the well studied WIMP annual modulation scenario \cite{freese1,freese2},
there is no spatial dependence.
The mass (and number) density is neither space nor time dependent,
and is estimated to be $\rho_0 \approx 0.3\ {\rm GeV}/{\rm cm}^3$ near the Earth.
All of the modulation arises from ${\bf v}_E(t)$,
understood simply as the time variation
of a Galilean boost through a Maxwellian velocity distribution.
The plasma dark matter scenario is distinctly different.
Far from the Earth, and for a fully ionised plasma,
the velocity distributions are (ideally) expected to be described by 
boosted Maxwellians in the Earth frame with
number densities:
\begin{eqnarray}
n_{e_d} = \frac{Z' \rho_0}{Z' m_{e_d} + m_{p_d}}, \ \ 
n_{p_d} = \frac{n_{e_d}}{Z'} \ .
\end{eqnarray}
However, this will not be the case in the vicinity of the Earth, where the detector is located.
As is evident from Figures~\ref{FigMoonB0} and \ref{FigVenusB0},
both the number density and the velocity distributions are expected
to display strong and non-trivial space and time dependence.
It is therefore necessary to time-average the rate over 
the detector path ${\bf x}(t)$, 
and this will introduce an important new source of modulation.
We will now describe this detector path.

\begin{figure}
 \centering
 \includegraphics[width=10.9cm,angle=0]{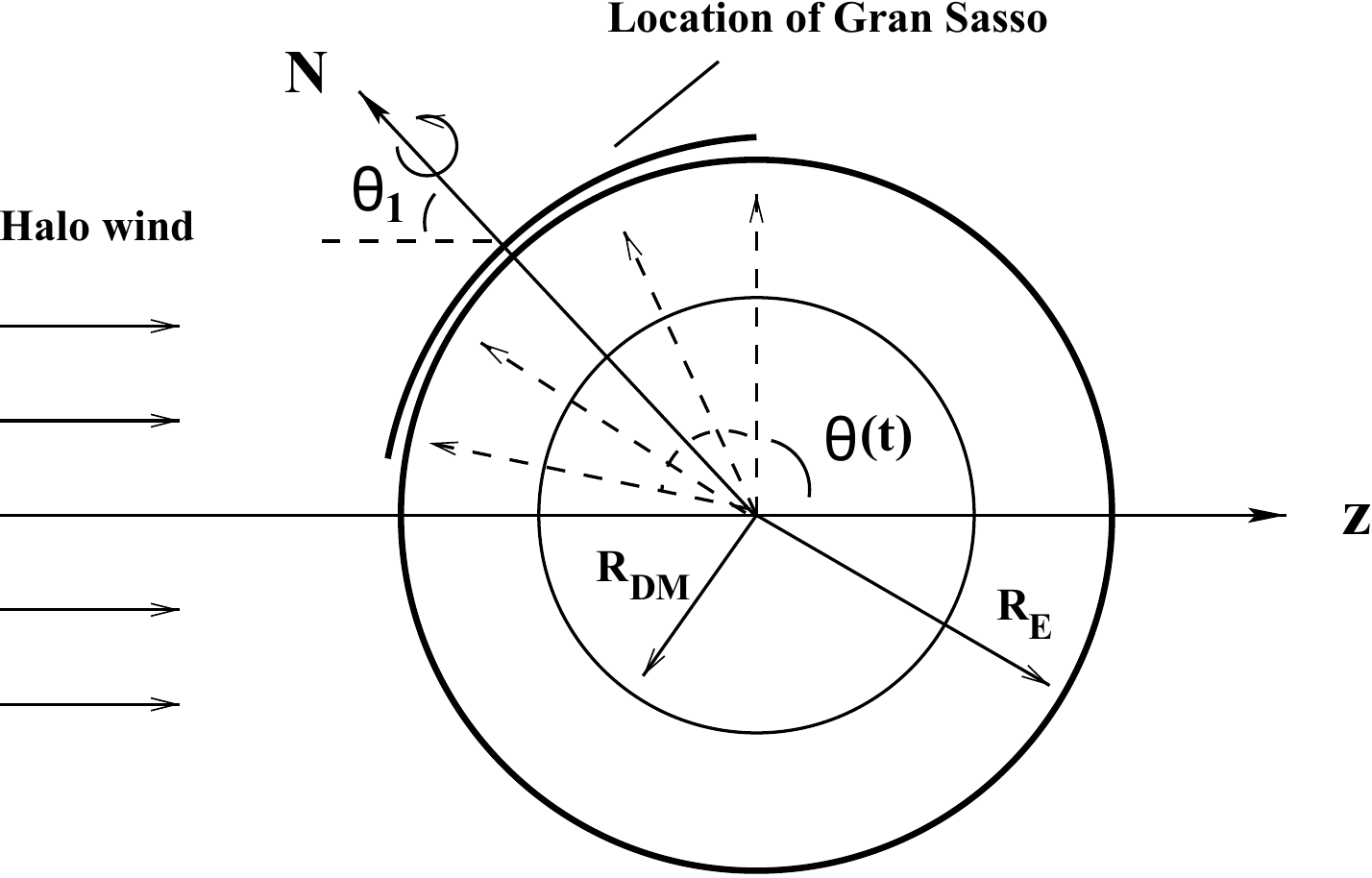}
 \caption{{\small 
 The relevant geometry ($r, \theta$ projection) of
 the dark halo wind interaction with the dark sphere (of radius $R_{DM}$)  within the Earth.
 The variation of the location of an example detector (Gran Sasso) due to the Earth's
 daily rotation is indicated.
 }}
 \label{geom}
\end{figure}

Consider the spherical coordinate system with its origin at the Earth's center 
and with ${\bf z}$ axis pointing in the direction
of the halo wind as shown in Figure~\ref{geom}. 
Assuming azimuthal symmetry around the ${\bf z}$ axis, 
the position of the detector is given in polar coordinates by ${\bf x}(t) = \left( R_E,\theta(t) \right)$,
where $\theta (t)$ is the angle between the direction of the halo wind and the zenith at the detector's location.
The time variation of the angle $\theta (t)$ is due
to the Earth's daily rotation and motion around the sun.
This angle has been evaluated previously \cite{fdiurnal} and is given by
\begin{eqnarray}
\cos\theta (t) = -\sin \theta_1(t) \cos\theta_{lat} \cos \left( \frac{2\pi t}{T_{day}}\right)  - \cos\theta_1(t)
\sin\theta_{lat} \ , 
\label{EqTheta}
\end{eqnarray}
where $T_{day} = 1$ {\it sidereal} day, and
the phase is such that $\theta (t)$ is maximised at $t=0$.
Here, $\theta_{lat}$ is the latitude of the detector's location,
which anticipates the important feature that {\it the measured rate and modulation will depend on the latitude of the detector}.
The parameter $\theta_1 (t)$ is the angle subtended by the direction of the Earth's motion through the halo  
with respect to the Earth's spin axis,
which varies during the year due to the Earth's motion around the sun: 
\begin{eqnarray}
\cos\theta_1(t) 
  \simeq  \cos \bar \theta_1 + y\left[ \cos \bar \theta_1 \cos \gamma  \sin \left( \frac{2\pi(t-T_1)}{{\rm year}} 
\right)
+ \sin \theta_{tilt} \sin \left( \frac{2\pi(t-T_2)}{{\rm year}}\right) \right] ,
\nonumber \\
& &
\label{EqTheta22}
\end{eqnarray}
where $\theta_{tilt} = 23.5^\circ$ is the angle between of the Earth's spin axis and the normal of the ecliptic plane,
$\gamma = 60^\circ$ is the angle between the normal of the ecliptic plane and the direction of the halo wind,
$T_1 = t_0  + 0.25 \ {\rm years} \simeq 244$ days, 
$T_2 \simeq 172$ days (northern summer solstice),
and $y = v_\oplus/v_\odot \approx 30/232 \approx 0.13$. 
Evidently the angle $\theta_1 (t)$ varies during the year with 
an average value of $\bar \theta_1 \simeq 43^\circ$,
a maximum of around $49^\circ$ on April $25^{th}$ (115 days), 
and a minimum of around $36^\circ$ six months later (297 days). 

\begin{figure}
 \centering
 \includegraphics[width=0.74\textwidth]{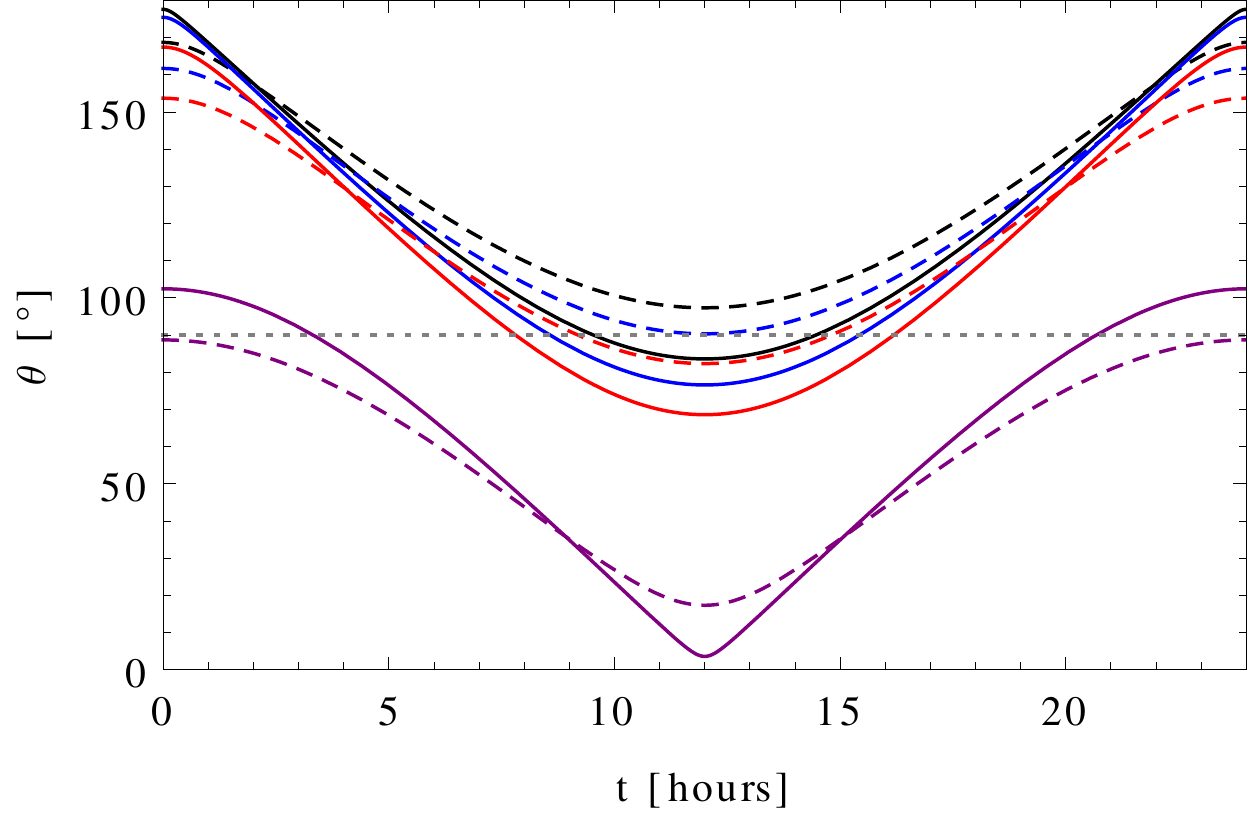}
 \caption{{\small
 The time variation in sidereal hours of $\theta (t)$, the angle between the direction of the halo wind and the zenith at the detector's location.
 The time variation of the angle $\theta (t)$ is due
 to the Earth's daily rotation and motion around the sun [Eq.~(\ref{EqTheta},\ref{EqTheta22})].
 Shown are results for four detector locations, for April 25 (solid-curves), October 25 (dashed-curves).
 The four detector locations are, from top to bottom: Gran Sasso (black curves), Kamioka (blue curves), Jin-Ping (red curves), and
 Stawell (purple curves).
 The $90^\circ$ line is also shown, which is the demarcation between the upwind and downwind regions.
 }}
\label{theta}
\end{figure}

In Figure~\ref{theta} we show the time variation of 
$\theta$ for four detector locations of interest:
Gran Sasso \cite{Bernabei2008yh,XENON100Exp} or Sanford \cite{Akerib2012ys},
Kamioka \cite{Abe2013tc},
China Jin-Ping \cite{Cao2014jsa},
and Stawell \cite{Shields2015wka}, 
which correspond to
$\theta_{lat} \approx 43^\circ,36^\circ,28^\circ,-37^\circ$,
respectively.
Also shown in Figure~\ref{theta} is the $\theta = 90^\circ$ line which is the demarcation between the
upwind and downwind regions. In this region, the $\rho, \ T, \ |v|$
quantities can vary significantly due to the tail shock feature, evident in Figures~\ref{FigMoonB0} and \ref{FigVenusB0}.
Interestingly the Gran Sasso laboratory spends most of its time in the near upwind region, while the Kamioka and Jin-Ping
laboratories loiter in the near downwind region. Thus, even the relatively small latitude difference between these laboratories
might be important, leading (potentially) to different dark matter interaction rates
for detectors at these locations.
Note that the Southern Hemisphere detector traverses a very different
path compared to the Northern Hemisphere detectors, and can potentially probe the downwind wake region for part of the day.

Let us summarise the origin of modulation signals in plasma dark matter models:
\begin{enumerate}
 \item Annual modulation with phase around June $2^{nd}$ (153 days)
 due to the variation of the Earth's speed relative to the dark matter halo.
 This variation not only acts to provide a Galilean boost with respect to the halo,
 as in the well studied WIMP scenario,
 but also to change the non-trival density and velocity distributions
 in the vicinity of the Earth.
 In the previous section we used MHD simulations to describe 
 some features of these distributions for two idealised scenarios,
 as shown in Figures~\ref{FigMoonB0} and \ref{FigVenusB0}.\footnote{There
are additional contributions to the annual modulation with phase June $2^{nd}$,
which we haven't considered, and may be important.
Among these are the variation of the physical properties of the dark sphere:
variation of the effective $R_{DM}$, the surface ionisation fraction,
``dark atmosphere'' interactions, etcetera.
}
 \item Annual modulation with phase around April $25^{th}$ (115 days)
 due to the variation of the Earth's spin axis relative to the wind direction.
 This effect changes the detector's daily path through the dark matter distribution
 according to Eq.~(\ref{EqTheta}), and can be the dominant source of annual modulation.
 \item Sidereal daily modulation due to the rotation of the Earth
 with respect to the direction of the plasma wind
 and the subsequent time-dependent position of the detector
 throughout the day, again according to Eq.~(\ref{EqTheta}).
 This is an extremely distinctive feature
 which can be probed with direct detection experiments.
 It is difficult to conceive any background which modulates with sidereal day.
\end{enumerate}
The latter two effects are, of course, generic predictions 
of any spatially dependent near-Earth dark matter density/velocity distribution.
Including, for example, models with dark matter subcomponents with sufficient interactions
with the ordinary matter within the Earth 
to be stopped (or at least impeded) \cite{Juan1,Juan2,denmark}.
In the plasma dark matter case the spatial dependence arises from the complex interaction between 
the dark plasma wind with the captured dark sphere within the Earth.
In the next section we will explore these modulation effects in an example model.

\section{Example: modulation of electron recoils in the mirror dark matter model}

So far we have only outlined the origin of
direct detection modulation signals in plasma dark matter models.
In this section we will explore more explicitly,
based on our MHD simulation results of section 4,
the range of possible annual and diurnal modulations.
The aim is to gain insight into where and how
direct detection experiments should be searching for plasma dark matter.
To facilitate this we will consider an explicit example: the mirror dark matter model.

The mirror model has fundamental plasma dark matter parameters
$m_{e_d} = m_e \simeq 0.511$~MeV, $m_{p_d} = m_{He} \simeq 3.76$~GeV, 
$Z' = 2$, and $\alpha_d = \alpha \simeq 1/137$.
Then, in the Milky Way,
$T\sim 0.35$~keV and $I/T\sim 0.16$,
so that the dark matter exists primarily in a plasma state.
Dark matter - ordinary matter interactions 
are due to the photon - dark photon kinetic mixing term Eq.~(\ref{kine}),
which induces Coulomb scattering 
of dark electrons (dark protons) against electrons (nuclei).
At $3.76$ GeV, the dark proton is just light 
enough so that nuclear recoil rates are strongly kinematically suppressed in current experiments.\footnote{
The mirror dark matter model can have heavier, mirror metal halo (sub)components, of known masses but
uncertain abundances. Several works (e.g. \cite{footdd1,footdd2}) have explored the possibility that these components might lead
to observable nuclear recoils in existing experiments.
However,
these studies used a very simplified picture for the halo distribution function, i.e.  without considering 
the modifications due to the interaction 
of the plasma wind with the dark sphere within the Earth as discussed here.
}
Of most interest, therefore, is dark electron - electron scattering.

Coulomb scattering of dark electrons off electrons ($e_d e \to e_d e$) is a spin-independent process  
with cross section:
\begin{eqnarray}
\frac{d\sigma}{dE_R} = \frac{\lambda}{E_R^2 v^2} ,
\label{cs}
\end{eqnarray}
where 
\begin{eqnarray}
\lambda \equiv \frac{2\pi \epsilon^2 \alpha^2}{m_e},
\end{eqnarray}
and $E_R$ is the recoil energy of the scattered electron, 
approximated as being free and at rest
relative to the incoming dark electron of speed $v$.
Naturally this approximation can only be valid for the loosely bound atomic electrons, i.e. those with binding energy much less
than $E_R$.

To proceed, we need to determine 
the local scattering rate as a function of position 
in the vicinity of the Earth.
To do this we have to evaluate Eq.~(\ref{f3}),
i.e. we have to integrate over the local velocity distribution.
Unfortunately our MHD simulations only tell us the local moments of this distribution.
Thus, without making a further assumption, we are stuck.
In order to continue,
{\it we will assume that the velocity distribution is everywhere locally 
given by a (boosted) Maxwellian}, i.e.
\begin{align}
 f_{e_d}({\bf v}) = \left( \frac{1}{\pi v_0^2} \right)^\frac32
  \exp\left( \frac{-({\bf v} - {\bf v}_B)^2}{v_0^2} \right),
\end{align}
where $v_0 = (2T/m_{e_d})^\frac12 \approx 11200\ (T/0.35\text{ keV})^\frac12$~km/s,
${\bf v}_B$ is the bulk velocity in the Earth (i.e. detector) frame,
and the space and time dependence is implied.
We do not expect this to be a good assumption in general.
Nevertheless, it will reproduce naive expectations
that the scattering rate scales positively with temperature and bulk velocity.
Due in a large part to this assumption we warn that our results
should be interpreted only as qualitative.
With this caveat acknowledged, 
the local differential rate Eq.~(\ref{f3}) can be evaluated 
for the (boosted) Maxwellian dark electron velocity distribution: 
\begin{align}
\frac{dR_e}{dE_R} = 
 \frac{N_T g_T n_{e_d} \lambda}{2E_R^2 |{\bf v}_B|} \ 
 \left[\text{erf}\left( \frac{v_{min}+|{\bf v}_B|}{v_0} \right) 
     - \text{erf}\left( \frac{v_{min}-|{\bf v}_B|}{v_0} \right) \right] 
\ .
\label{r68}
\end{align}
Here $N_T$ is the number of target particles
(e.g. NaI pairs for DAMA, and Xe for the xenon experiments),
$g_T$ is the effective number of ``free'' electrons 
(binding energy $\lesssim 1$~keV) per target particle
($g_{\text{NaI}}\approx 54$, $g_{\text{Xe}}\approx 44$),
and $v_{min} \approx 26500\ (E_R/2 \text{ keV})^\frac12$~km/s [from Eq.~(\ref{v})].\footnote{The 
differential event rate evaluates (for a NaI detector) to:
{\scriptsize
\setlength{\abovedisplayskip}{1pt}
\setlength{\belowdisplayskip}{4pt}
\begin{align}
\frac{dR}{dE_R} \approx 0.6 
  \left(\frac{n_{e_d}}{0.16\text{ cm}^{-3}} \right)
  \left(\frac{\epsilon}{10^{-9}}\right)^2
  \left(\frac{2\text{ keV}}{E_R}\right)^2
  \left(\frac{0.35\text{ keV}}{T}\right)^{\frac12}
  \text{exp}\left[ -\frac{2}{0.35}\left( \frac{E_R/2\text{keV}}{T/0.35\text{keV}} - 1 \right)\right]\text{ cpd/kg/keV}.
  \nonumber
\end{align}
}%
This can be compared with the rough limit from DAMA that the differential rate 
should be less than about 0.25~cpd/kg/keV at $E_R\simeq 2$~keV \cite{dam9}.
Evidently $\epsilon$ in the range $10^{-9} - 10^{-10}$
is being probed in direct detection experiments via electron recoils in the mirror model, which
coincides with the range of interest for small scale structure \cite{sph,mirror1,rich8}.
}

\begin{figure}
 \centering
 \includegraphics[width=0.74\textwidth]{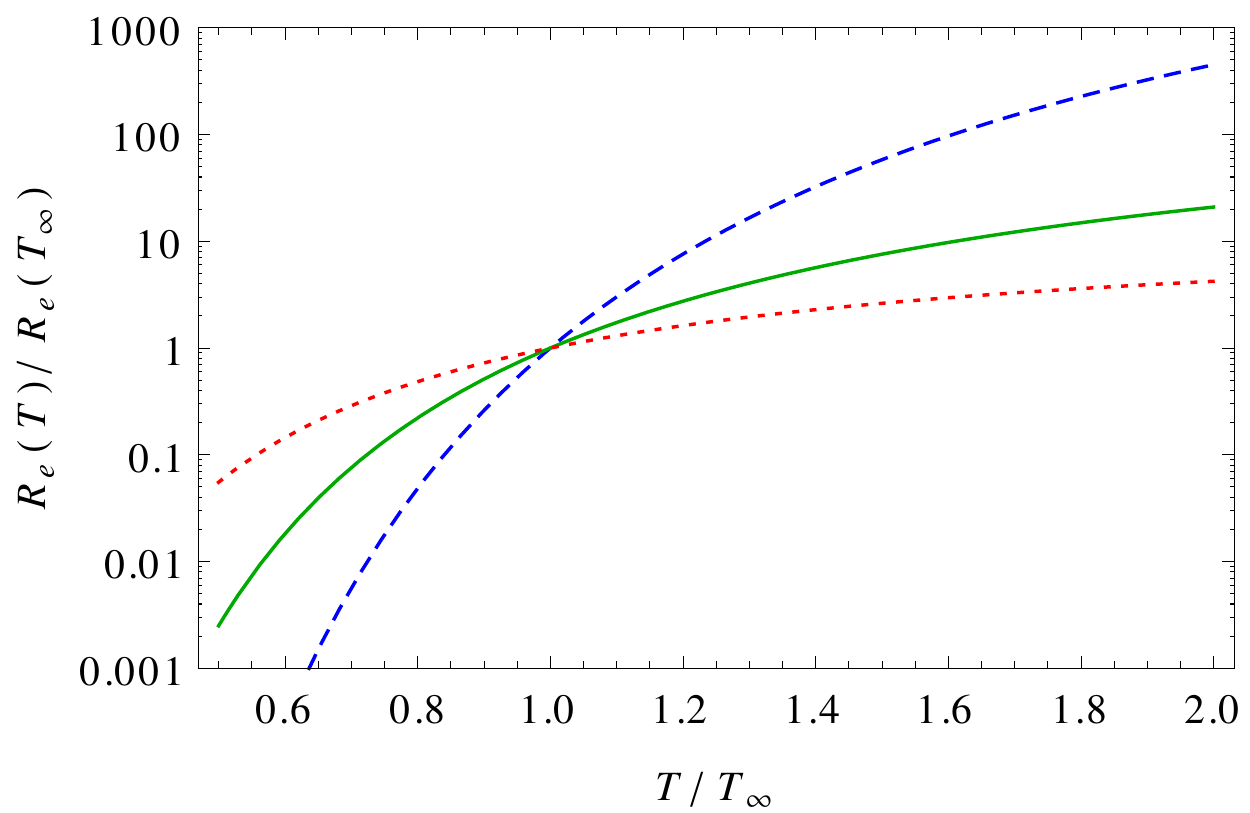}
 \vskip -0.2cm
 \caption{{\small
 Rate dependence on temperature for 
 $T_\infty=0.17,0.35,0.73$~keV
 (or $c_s=140,200,290$~km/s)
 as blue dashed, green solid, red dotted, respectively.
 A low energy threshold of $E_t = 2$ keV is assumed.
}}
 \label{FigTFunc}
\vskip 0.5cm
\end{figure}

For electron recoils in the mirror dark matter model, 
$|{\bf v}_B| \ll v_{0}$, and in the limit $|{\bf v}_B|/v_{0} \to 0$ Eq.~(\ref{r68}) 
can be integrated 
from a threshold energy, $E_t$,
to give:
\begin{align}
 R_e = N_T g_T n_{e_d} \lambda \left(\frac{2 m_{e_d}}{\pi T}\right)^\frac12
 \left(
   \frac{e^{-\frac{E_{t}}{T}}}{E_t}  - \frac{\Gamma\left[0,\frac{E_t}{T} \right]}{T}
 \right), \label{EqTotRate}
\end{align}
where $\Gamma[0,z]$ is the upper incomplete Gamma function.
Corrections due to non-zero ${\bf v}_B$ are $\mathcal{O}(|{\bf v}_B|^2/v_0^2)$
and remain below one per cent for all cases considered.
We note here that this rate is dominated by low energy recoils,
and is therefore very sensitive to the lower limit of integration, $E_t$.
For $E_t = 2$ keV (DAMA value),
it is also a rather sensitive function of $T$,
since $E_R>2$~keV requires $v_{e_d}>26500$~km/s,
in the tail of the dark electron velocity distribution.
In Figure~\ref{FigTFunc} we illustrate this sensitivity
for the range of $T_\infty$ we consider.
These sensitivities are further reasons to interpret our results only qualitatively.

The local differential rate, Eq.~(\ref{r68}), 
is a function of the local $T$, $\rho$, and $|{\bf v}_B|$
(the rate depends on $\rho$ via its dependence on $n_{e_d}$).
In particular these space and time dependent quantities were obtained 
in section~4, as shown in Figures~\ref{FigMoonB0} and \ref{FigVenusB0}.
Fixing $v_{rot} = 220$ km/s to give the time-dependent dark plasma wind velocity 
in the Earth frame [Eq.~(\ref{EqvE})],
the space and time dependence of the differential rate throughout the year will
depend only on the plasma sound speed $c_s$, or equivalently 
$T_\infty~=~0.35\ [c_s/(200\text{ km/s})]^2$~keV. 
Then for a given detector latitude $\theta_{lat}$
and dark sphere size $R_{DM}$ 
it is possible to determine the quantities of interest:
the rate as a function of time of year averaged over the day
(the annual modulation),
and;
the rate as a function of time of sidereal day averaged over the year
(the diurnal modulation).
This average includes the variation throughout the year 
of dark plasma wind Mach number
and position of the detector
according to Eqs.~(\ref{EqvE}) and (\ref{EqTheta}), respectively.

\begin{figure}
 \hspace{0.0cm}
 \makebox[\textwidth][c]{
 \includegraphics[width=1.15\textwidth]{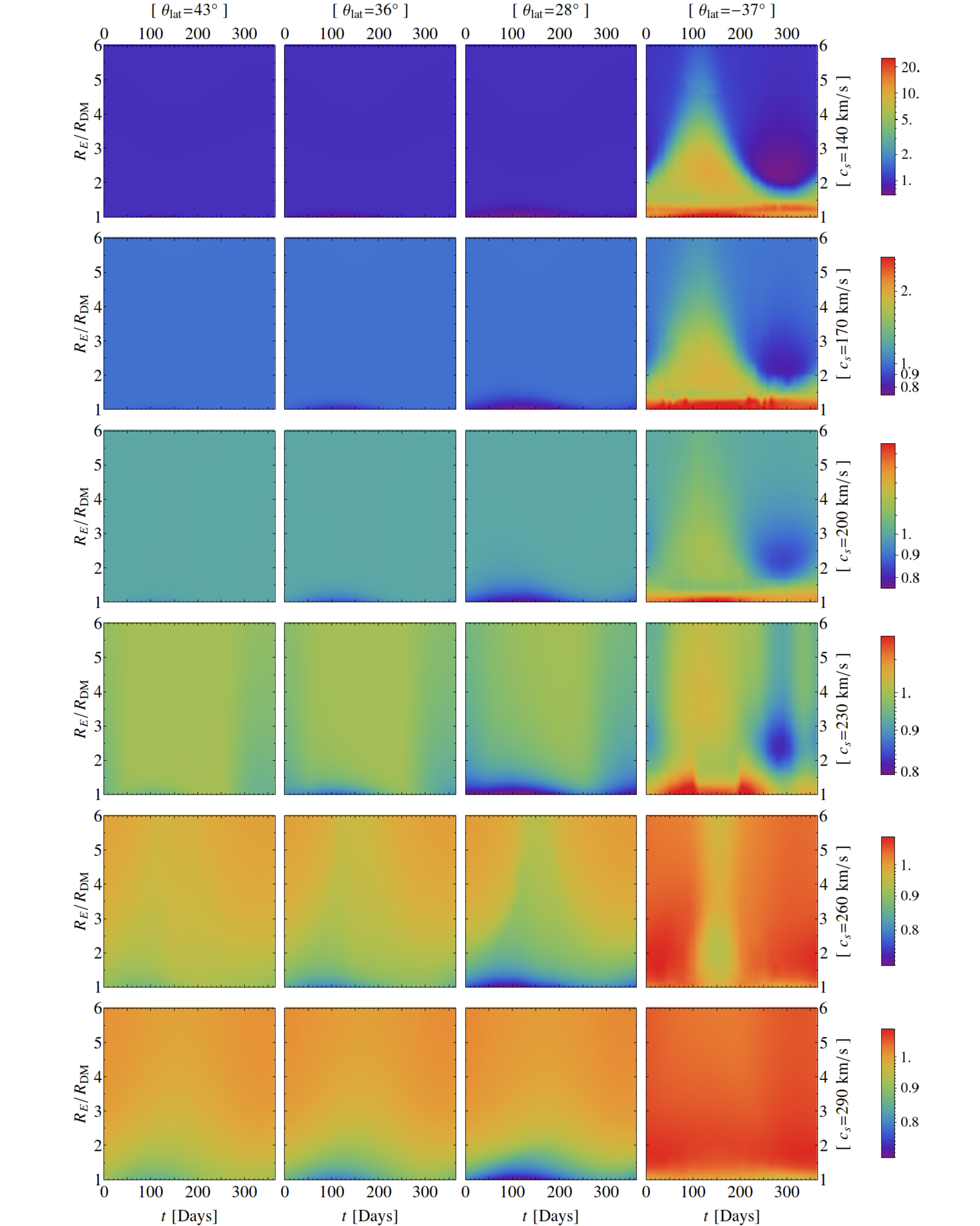}}
 \caption{{\small Moon-like ($B=0$) annual modulation: 
 $R_e/R_e^\infty$ as a function of time of year
 plotted for example detector locations [Gran Sasso, Kamioka, Jin-Ping, and Stawell] (columns) and sound speeds (rows).
 }}
 \label{FigAnnMoonB0}
\end{figure}

\begin{figure}
 \hspace{0.0cm}
 \makebox[\textwidth][c]{
 \includegraphics[width=1.15\textwidth]{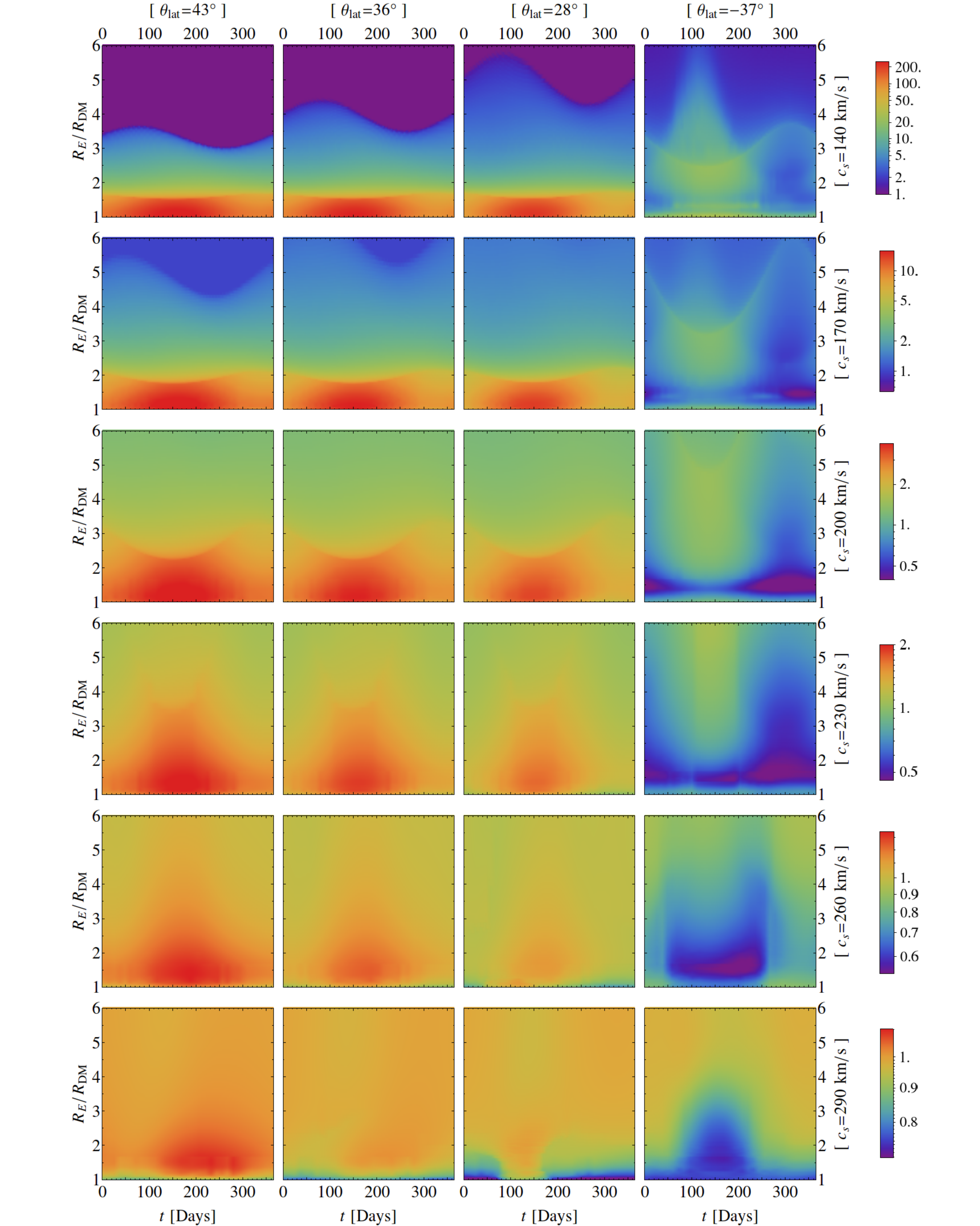}}
 \caption{{\small Venus-like ($B=0$) annual modulation: 
 $R_e/R_e^\infty$ as a function of time of year
 plotted for example detector locations [Gran Sasso, Kamioka, Jin-Ping, and Stawell] (columns) and sound speeds (rows).
 }}
 \label{FigAnnVenusB0}
\end{figure}

\begin{figure}
 \hspace{0.6cm}
 \makebox[\textwidth][c]{
 \includegraphics[width=1.15\textwidth]{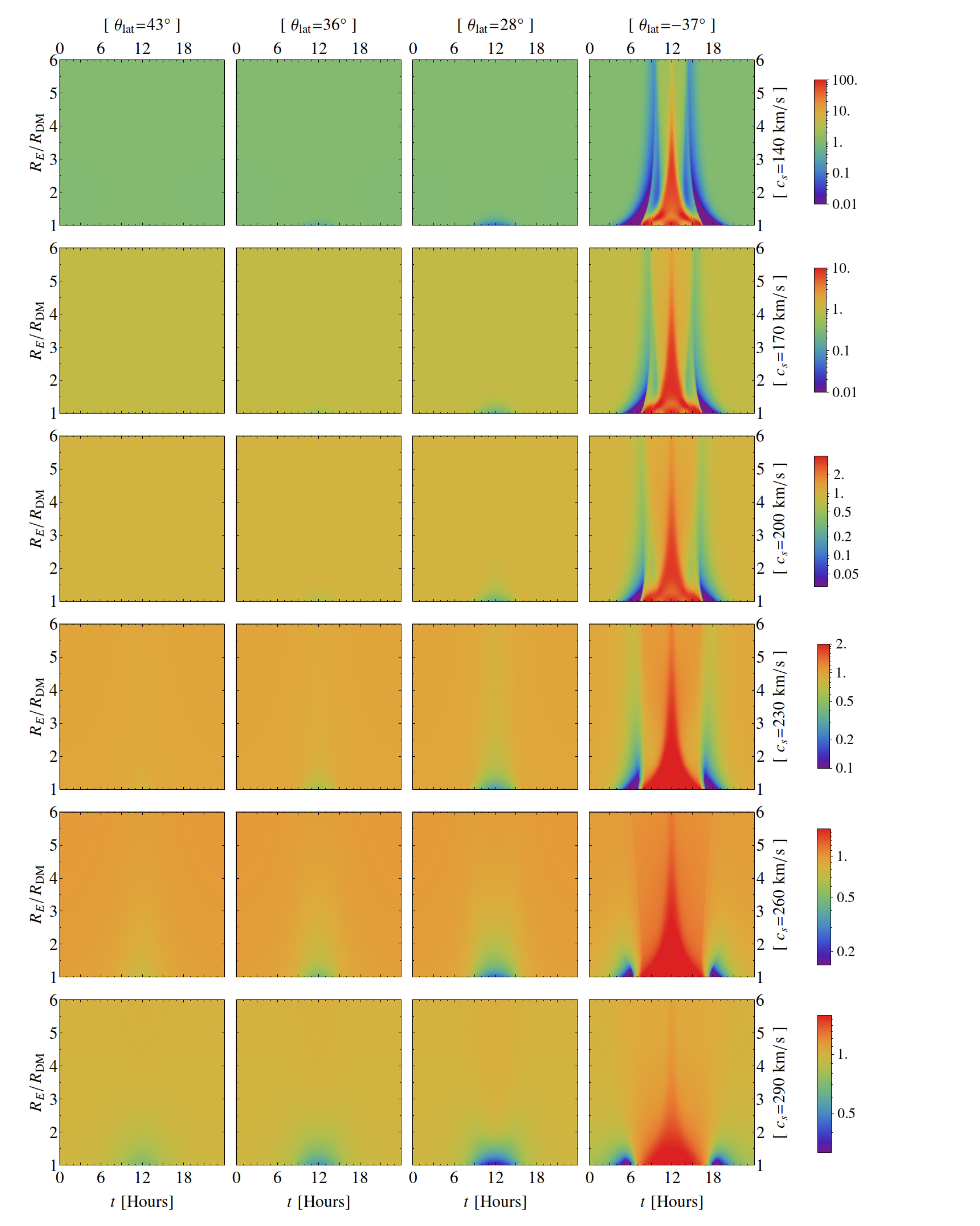}}
 \caption{{\small Moon-like ($B=0$) diurnal modulation: 
  $R_e/R_e^\infty$ as a function of sidereal hours 
  plotted for example detector locations [Gran Sasso, Kamioka, Jin-Ping, and Stawell] (columns) and sound speeds (rows).
 }}
 \label{FigdayMoonB0}
\end{figure}

\begin{figure}
 \hspace{0.6cm}
 \makebox[\textwidth][c]{
 \includegraphics[width=1.15\textwidth]{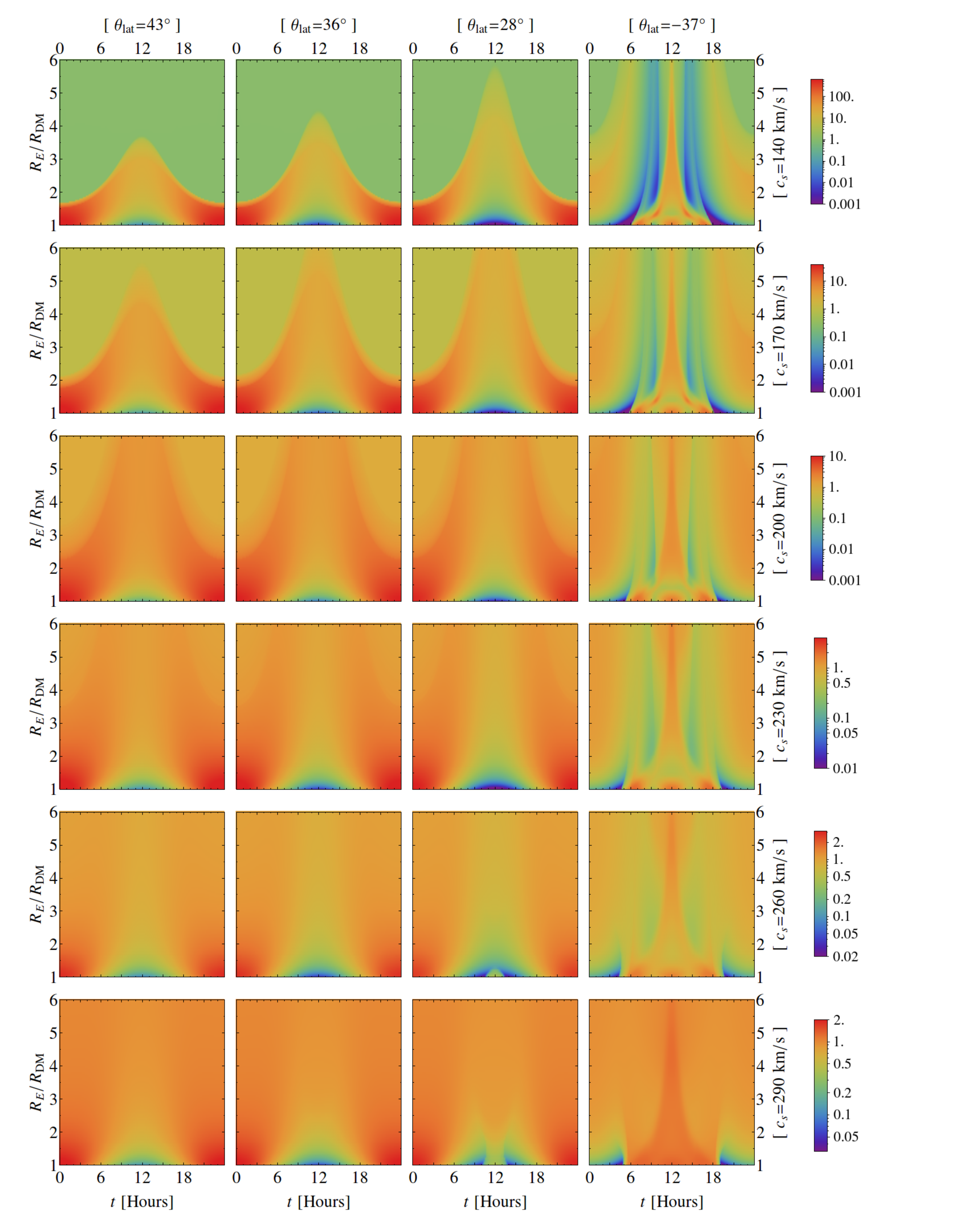}}
 \caption{{\small Venus-like ($B=0$) diurnal modulation: 
  $R_e/R_e^\infty$ as a function of sidereal hours 
  plotted for example detector locations [Gran Sasso, Kamioka, Jin-Ping, and Stawell] (columns) and sound speeds (rows).
 }}
 \label{FigDayVenusB0}
\end{figure}

For the numerical work, we set $E_t = 2$ keV (current DAMA threshold) and
considered an idealised detector with 100\% detection efficiency and perfect resolution.
We give our results in Figures~\ref{FigAnnMoonB0}--\ref{FigDayVenusB0}
under each of the scenarios previously considered, 
i.e. Moon-like/Venus-like dark sphere 
with unmagnetised plasma wind.
We consider sound speeds $c_s=140$--290~km/s,
encompassing the supersonic to subsonic dark plasma wind regimes,
and $\theta_{lat}=43^\circ,36^\circ,28^\circ,-37^\circ$,
which correspond to detectors at 
Gran Sasso \cite{Bernabei2008yh,XENON100Exp} or Sanford \cite{Akerib2012ys},
Kamioka \cite{Abe2013tc},
China Jin-Ping \cite{Cao2014jsa},
and Stawell \cite{Shields2015wka}, respectively.
We leave the size of the Earth with respect to the dark sphere, 
$R_E/R_{DM}$, as a free parameter which is assumed 
to remain constant throughout the averaging procedure.
Strictly, the particle physics should dictate
the nature and size of the dark sphere.
Indeed, for mirror dark matter,
the procedure described in section 3 suggests
a Moon-like scenario with $R_E/R_{DM}\approx 1$--1.5 .
Still, it is possible that effects such as surface ionisation
and dark atmospheric stripping significantly change this picture.
Thus it is sensible to consider each scenario and a range of dark sphere radii,
and this agnosticism anyway coincides with our aim to explore the range of
modulation possibilities in plasma dark matter models in general.

In the next section we will make some qualitative observations from these results
and deduce the implications for direct detection experiments.

\section{Plasma dark matter direct detection: Implications}

In the previous section we presented results for the annual and diurnal modulation
of electron recoils in the  mirror dark mater model.
Here we gather some comments on these results and discuss the current experimental situation.

The behaviour exhibited in Figures~\ref{FigAnnMoonB0}--\ref{FigDayVenusB0}
is clearly quite diverse, and we make the following observations:
\begin{enumerate}

\item The annual modulation fraction can be large, even $>90\%$.

\item There are situations where the modulation is approximately sinusoidal,
but this is not the general case.
Interplay with various shocks may produce sharp transitions in the rate.

\item The two annual modulation contributions, 
with phases of 153 days and 115 days, 
can be seen by eye, 
e.g. Moon-like $c_s\ge 260$~km/s (Figure~\ref{FigAnnMoonB0})
or Venus-like $c_s\le 170$~km/s (Figure~\ref{FigAnnVenusB0}).
The contributions can be different sign and either might dominate.

\item The dependence on $\theta_{lat}$ is obvious.
In particular, the Stawell detector gives very different results
since it probes the downwind wake region.
Still, even between Northern Hemisphere detectors, 
significant changes in the modulation can be observed.
For example, in the Moon-like case with $c_s$ in the intermediate region ($c_s\approx 230$~km/s)
with $R_E/R_{DM}\lesssim 2$ we observe 
a change in the effective sign between Northern Hemisphere detectors.
This is due to the interplay of the wind speed pushing the tail shock back around 155 days,
and the detector moving further into the shock at 115 days.

\item It is possible to see modulation effects in a detector at one latitude
which would escape detection in an identical detector at a different latitude.

\item The diurnal modulation fraction can be large,
typically of order the annual modulation, or larger.
It is in general not sinusoidal and can display sharp transitions.

\item The azimuthal symmetry (in the unmagnetised case) implies that diurnal modulation
is symmetric about $t=12$ hours (with our phase convention).
This motivates combining data from $0 \le t/{\rm hours} \le 12$
with data from $24 \ge t/{\rm hours} \ge 12$.
As well, the average rate during the middle half of the day (6--18 hours)
often differs markedly from the average rate during than the other half.
This motivates a far/near ratio measurement,
\begin{equation}
 R_{{\rm far/near}} = 
 \frac{ R( 6 \le t/{\rm hours} \le 18) } 
      { R( 0 \le t/{\rm hours} \le 6 \ \cup \ 18 \le t/{\rm hours} \le 24) } \ ,
 \label{EqFarNear}
\end{equation}
looking for deviations from unity.

\end{enumerate}

Nuclear recoils arising from dark proton scattering are, in principle, also very interesting. 
Although we haven't given any results for nuclear recoils, qualitatively they are expected
to follow a similar pattern to the electron recoil results of 
Figures~\ref{FigAnnMoonB0}--\ref{FigDayVenusB0}.
Indeed, the nuclear recoil rate has a similar form to Eq.~(\ref{r68}),
and in particular, the rate depends on the same variables
$\rho,\ T, \ |{\bf v}_B|$ (in the single fluid approximation). 
The principle difference is the nuclear recoil kinematics, 
which depend on the mass of the dark proton.
In general we expect nuclear recoils to show similar sensitivity to variation in $\rho$, 
comparable or smaller sensitivity to variation in $T$ (depending on the mass of the dark proton),
and more sensitivity to variation in $|{\bf v}_B|$
(since $|{\bf v}_B|/v_0$ is larger by a factor $\sqrt{m_{p_d}/m_{e_d}}$).

Let us here add somewhat of a disclaimer.
We have sketched what we believe to be a sound general picture
of modulation effects in plasma dark matter models.
However, there is great difficulty involved in modelling 
the interaction of the dark plasma wind with the captured dark sphere.
We have made a first attempt at a consistent description using MHD simulations.
Obviously there are shortcomings.
Notably, we have only studied two idealised dark sphere scenarios in the special unmagnetised case,
and we have also made the questionable assumption of locally Maxwellian distributions
in order to explore the modulation signals for an example model.
Quantitative results for the actual realised case may be very different.
Nonetheless, we believe that the above qualitative observations should still hold.

We have yet to discuss the current experimental situation with regard to dark matter direct detection.
A variety of experiments, employing different techniques,
are probing dark matter interactions with nuclei and electrons.
Stringent limits on dark matter nuclear recoils have been found, 
with XENON100 \cite{xenon100}, LUX \cite{lux}, CRESST-II \cite{cresst}, 
and CDMS \cite{cdms} among the most sensitive.
By contrast, a positive hint for dark matter interactions has been obtained by
the DAMA and DAMA/LIBRA experiments in the Gran Sasso Laboratory (latitude: $43^\circ$ N) \cite{dama1a,dama1b,dama2a,dama2b,dama2c}.
The DAMA experiments were designed to search for dark matter via the annual modulation signal and indeed  
such a modulation (with phase: $t_0 = 144 \pm 7$ days) 
was observed in their measured event rate at around  $\sim 9\sigma$ C.L..
The DAMA and DAMA/LIBRA experiments feature
a sodium iodine target with sensitivity to both nuclear and electron recoils in the keV recoil energy range.
The stringent limits on nuclear recoils obtained by other experiments (as mentioned above) appears to indicate  
that electron recoils is the most likely dark matter option.

There are only a few experiments with sufficient sensitivity to 
probe electron recoils as the source of the DAMA annual modulation signal.
At the present time, three such experiments have published results:  
CoGeNT, XENON100, and XMASS, all of which have some, albeit statistically weak, evidence
for an annually modulated event rate.
Consider first the CoGeNT experiment. This experiment involves p-type point contact germanium detectors
operating in the Soudan Underground Laboratory (latitude: $48^\circ$ N). 
Analysis of three years of data found evidence for an annual modulation at $2.2 \sigma$ C.L. 
with phase consistent with that of DAMA \cite{cogent}.
The XENON100 experiment, located at Gran Sasso, 
recently analysed data collected over a 13 month period,
observing an annually modulated electron recoil event rate at $2.8 \sigma$ C.L.
with phase consistent with that of DAMA \cite{xenon100electron}.
The XENON100 experiment also obtained strong limits 
on the average electron recoil event rate, 
thereby suggesting that dark matter
interactions with electrons could only be the source of the DAMA annual modulation 
if the modulation fraction was large: $\gtrsim 50\%$
\cite{xenon100electron,xenon100electron2}.
Most recently, the XMASS experiment at Kamioka Observatory (latitude: $36^\circ$ N), 
also utilising a xenon target,
has searched for dark matter - electron interactions \cite{xmass}. 
Their data shows a possible hint of annual modulation with opposite sign to that of DAMA
(i.e. approximately six months out of phase).
Naturally it is difficult to directly compare DAMA's annual modulation signal 
with the results of these other experiments, as they differ in their
recoil energy range, energy resolution, and low energy cutoff. 
The CoGeNT and XMASS experiments are also at different latitudes.

\begin{figure}
 \centering
 \includegraphics[width=7.3cm,angle=270]{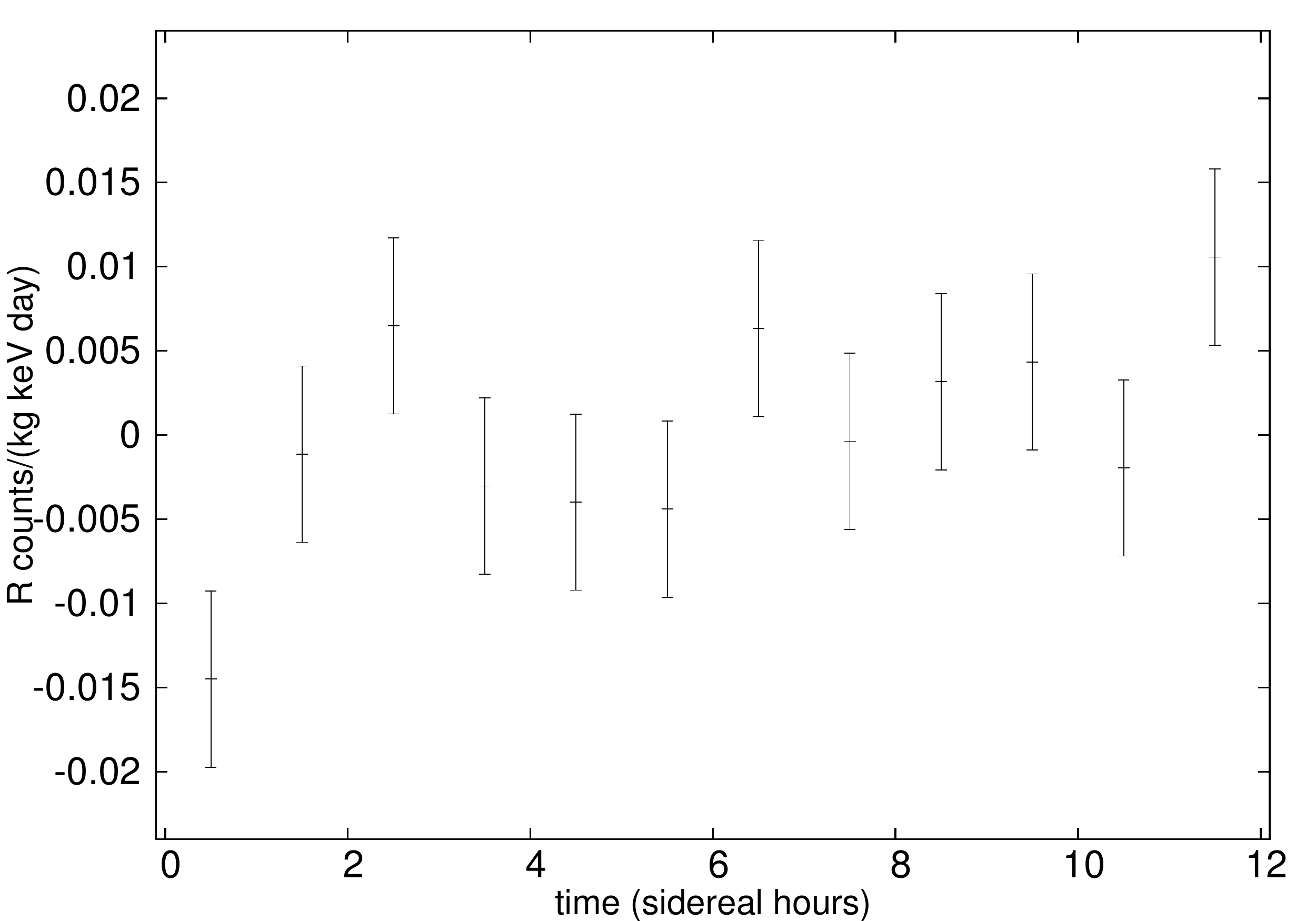}
 \caption{{\small
The DAMA \cite{ber9z} measured rate: $R - \langle R \rangle$ versus sidereal time, where the
data has been  replotted here with $24 \ge t \ge 12$ hours  combined with 
$0 \le t \le 12$ hours.
}}
\label{diurnal}
\end{figure}

Of these experiments, only DAMA has given results 
for their event rate binned into 24 sideral hours (i.e. diurnal modulation).  
Taking our phase convention, where $t=0$ is the time of day when $\theta$ is maximised, 
and motivated by azimuthal symmetry,
it is sensible to combine the 
data from $0 \le t/{\rm hours} \le 12$
with data from $24 \ge t/{\rm hours} \ge 12$.
We plot the data \cite{ber9z} combined in this way in Figure~\ref{diurnal}. 
The figure does show some modest evidence for a rising event rate toward $t = 12$ hours.
The far/near ratio Eq.~(\ref{EqFarNear}) can be evaluated as:
\begin{eqnarray}
R_{{\rm far/near}} = 1.0072\pm 0.0031 \ .
\end{eqnarray}
That is, $R_{{\rm far/near}}$ is different from unity at approximately 2.3$\sigma$ C.L..

The current experimental situation is rather intriguing, 
especially when viewed in the context of plasma dark matter.
Indeed, plasma dark matter appears to have the potential to resolve
the diverse results of the different experiments. 
In particular, our analysis leaves open the interesting possibility 
that the DAMA annual modulation signal
might be due to electron recoils (or even a combination of electron and nuclear recoils). 
This modulation fraction can be large, thus potentially satisfying the constraints
on electron recoils from XENON100. 
Similarly, constraints on nuclear recoils (such as those in \cite{xenon100,lux,cresst,cdms}) are considerably 
weakened if the modulation fraction is large.
Also, the results of XMASS might not be inconsistent with DAMA given the difference in latitude between the locations of
these two experiments. 

Clearly further work is required to clarify this situation. 
More experiments could analyse their data for possible dark matter interactions with electrons,
in addition to nuclear recoils.
Diurnal modulation in addition to annual  
modulation should be searched for.
Experiments at different latitudes are important, and more experiments in the Southern 
Hemisphere would be helpful.

\section{Conclusions}

Dark matter may have non-trivial particle properties leading to all sorts of interesting effects on small scales.
The particular situation studied here is that
dark matter in spiral galaxies like the Milky Way takes the form of a dark plasma.
Hidden sector dark matter charged under an unbroken $U(1)'$ gauge interaction provides a 
simple and well defined particle physics model
realising this possibility.
The assumed $U(1)'$ neutrality of the Universe then implies (at least) two oppositely charged dark matter components
with self-interactions mediated via a massless ``dark photon''  (the $U(1)'$ gauge boson).
We considered the simplest case of two such dark matter components, 
the ``dark electron''  and the ``dark proton'', with $m_{e_d}\le m_{p_d}$.

Various astrophysical and cosmological aspects of this type of dark matter have been explored in the literature previously,
but there have been relatively few attempts to understand the implications for direct detection experiments.
This seems to be particularly relevant at the present time in view of the rapidly progressing experimental activity in 
the field of dark matter direct detection.
Moreover, plasma dark matter is quite unique 
in that it can potentially lead to both nuclear and electron recoils
in the keV energy range;
this is because energy equipartition implies a potentially large dark electron velocity dispersion,
and $U(1)'$ neutrality prevents dark electrons from escaping the galaxy.
In fact, previous work has speculated that plasma dark matter might possibly be able to explain the DAMA annual 
modulation signal via electron recoils, as the constraints on electron recoils from other experiments are generally
much weaker than those for nuclear recoils.

To properly examine this idea, and the implications for direct detection experiments more generally, requires
a detailed description of the plasma dark matter density and velocity
distribution in the vicinity of the Earth.
This is a rather complex problem as any assumed interaction with ordinary matter will inevitably lead to
dark matter being captured by the Earth, forming an approximate ``dark sphere'' within.
This dark sphere provides an obstacle to the halo dark matter wind, the nature of which 
depends on whether the captured dark matter is largely neutral or ionised.
We considered these two
limiting cases, referred to as ``Moon-like'' or ``Venus-like,'' making use of analogy 
with the solar wind interactions with the Moon and Venus. 
We studied these limiting cases using single fluid magnetohydrodynamic equations.

We numerically solved the magnetohydrodynamic equations 
to obtain the space and time dependent 
dark plasma density, temperature, and bulk velocity
in the vicinity of the Earth.
We identified two distinct sources
of annual modulation:
the first arises from the variation of the Earth's speed relative to the dark matter halo, and;
the second arises from the variation of the Earth's spin axis relative to the wind direction.
While both effects are due to the Earth's orbital motion around the Sun, their phases are different:
June $2^{nd}$ versus April $25^{th}$. 
In addition, the variation of the location of a given detector
relative to the wind direction due to the Earth's daily rotation leads to a diurnal modulation 
(i.e. with period of one sidereal day).
The latter two modulation effects are a direct consequence
of the spatially dependent near-Earth dark matter density and velocity distributions,
and are expected to be an important consideration 
in general self-interacting dark matter models capable of giving a direct detection signal.
Importantly, they imply latitudinal dependence of the measured event rate.

In order to make predictions for direct detection experiments, a
kinetic description of the plasma
dark electron and dark proton components is required.
This is a challenging and unsolved problem.
To make progress, we modelled the velocity distribution
locally in terms of a Maxwellian distribution.
Although this is rather unsatisfactory, it
is hoped that such a description will provide useful insight.
We considered mirror dark matter as an example, 
and evaluated the annual and diurnal modulations,
focusing on the distinctive electron recoil interaction.
Several relevant qualitative observations were made from the results.

Plasma dark matter is very different from e.g. weakly interacting dark matter.
Large annual and diurnal modulations can arise.
These modulations need not be sinusoidal and may contain sharp features.
Moreover,
the spatial dependence of the local event rate in the vicinity of the Earth
implies that experiments at different latitudes will not necessarily find the same thing (even qualitatively).
This is especially true for a Southern Hemisphere detector,
but is even true for varying latitudes in the Northern Hemisphere.
The analysis presented here 
leaves open the interesting possilibility that the DAMA annual modulation signal
might be due primarily to electron recoils (or even a combination of electron and nuclear recoils). 
The modulation fraction can be large, thus potentially satisfying constraints
from other experiments. 
Furthermore, the results of XMASS might not be inconsistent with DAMA given the difference in latitude between the locations of
these two experiments. 
Much more experimental activity is required. A greater emphasis on electron recoils would be helpful
and we encourage all experiments to present results for diurnal variation.

\vskip 1cm

\noindent 
{\large \bf Acknowledgments}
\vskip 0.2cm
\noindent
This work was supported by the Australian Research Council.
JDC would like to thank Chris Nolan for the suggestion of using \textsc{Pluto}
for the numerical work.

\bibliographystyle{JHEP.bst}
\bibliography{references}


\end{document}